\documentclass[prl,showpacs,superscriptaddress]{revtex4} 

\usepackage{amsmath,amssymb}
\usepackage{verbatim}
\usepackage{graphicx}
\usepackage{hyperref}
\usepackage{color}
\usepackage{footnote}
\usepackage{enumitem}

\DeclareFontFamily{OT1}{rsfs}{}
\DeclareFontShape{OT1}{rsfs}{m}{n}{ <-7> rsfs5 <7-10> rsfs7 <10->rsfs10}{} 
\DeclareMathAlphabet{\mycal}{OT1}{rsfs}{m}{n}

\newcommand{\be}[1]{ \begin{equation}\label{#1} }
\newcommand{\ee}{\end{equation}}
\newcommand{\bea}[1]{\begin{eqnarray}\label{#1} }
\newcommand{\eea}{\end{eqnarray}}

\newcommand{\gam}{\gamma_{ij}^{(1)}}

\newcommand{\eq}[2]{\begin{equation} #1 \label{#2} \end{equation}}

\newcommand{\al}{\alpha}
\newcommand{\ga}{\gamma}
\newcommand{\de}{\delta}

\newcommand{\Om}{\Omega}

\DeclareMathOperator{\extdm}{d}
\newcommand{\extd}{\extdm \!}

\newcommand{\ms}[1]{\textrm{\tiny $#1$}}

\newcommand{\FO}{\ms{(1)}}

\newcommand{\kvth}{\xi^{(3)}}
\newcommand{\kvfo}{\xi^{(4)}}
\newcommand{\kvfi}{\xi^{(5)}}

\newcounter{rowcount}
\usepackage{etoolbox}
\makeatletter
\patchcmd{\chapter}{\if@openright\cleardoublepage\else\clearpage\fi}{}{}{}
\makeatother

\begin{document}

\title{Asymptotic Symmetry Algebra of Conformal Gravity}


\author{Maria Irakleidou}
\email{irakleidou@hep.itp.tuwien.ac.at}
\affiliation{Institute for Theoretical Physics, Vienna University of Technology, Wiedner Hauptstrasse 8--10/136, A-1040 Vienna, Austria}

\author{Iva Lovrekovic}
\email{lovrekovic@hep.itp.tuwien.ac.at}
\affiliation{Institute for Theoretical Physics, Vienna University of Technology, Wiedner Hauptstrasse 8--10/136, A-1040 Vienna, Austria}


\date{\today}

\preprint{TUW--13--xx}

\begin{abstract} 
We compute asymptotic symmetry algebras of conformal gravity. Due to more general boundary conditions allowed in conformal gravity in comparison to those in Einstein gravity, we can classify the corresponding algebras. The highest algebra for non-trivial boundary conditions is five dimensional and it leads to global geon solution with non-vanishing charges.
\end{abstract}


\maketitle

\tableofcontents


\section{Introduction}
Conformal gravity (CG) is higher derivative theory of gravity that has recently obtained many interests since it can reproduce the solutions obtained by Einstein gravity (EG) \cite{Maldacena:2011mk}, which is one of the main requirements for the effective gravity theory.

Its advantage is that it is two loop renormalizable while EG is not, however it contains ghosts while EG is ghost free.
%
%
From theoretical aspects CG has been studied by 't Hooft in series of articles \cite{Hooft:2009ms,Hooft:2010ac,Hooft:2010nc}. In \cite{Hooft:2014daa}, he considers conformal symmetry to play a fundamental role for understanding the physics at the Planck energy scale. 
It arises from the twistor string theory \cite{Berkovits:2004jj} and as a counterterm from five dimensional EG. 
On the phenomenological grounds 
it  has been studied by Mannheim to explain the galactic rotation curves without the addition of the dark matter \cite{Mannheim:2012qw, 
Mannheim:2010xw}. 
Furthermore, it was proven in  \cite{Grumiller:2013mxa} that the analogous, linear, term which in two dimensional toy model provides additional matter, is allowed to appear in CG.
In four dimensional CG, that term leads to the response function that appears in CG beside the standard EG response function, which is Brown--York stress energy tensor.
This linear term was included taking into account the most general boundary condition, as explained in \cite{Grumiller:2013mxa}. 
Asymptotic symmetry algebra (ASA) of CG for vanishing of the linear term is conformal algebra, which is known to be ASA of Einstein gravity.  The equations of motion of CG allow for linear term in the expansion of the metric, while equations of motion of EG do not. Linear term can be further restricted with the choice of its components. Taking the  conformal algebra  as ASA for trivial boundary conditions ($\text{linear term}=0$) we show that the boundary conditions can be classified according to the subalgebras when $\text{linear term}\neq0$.

The response functions define boundary charges in the $AdS/CFT$ framework which define the asymptotic symmetry algebra (ASA) at the boundary. We consider the richness of the ASA depending on the restrictions of the imposed boundary conditions. 
In general, one is interested to study asymptotic symmetries for three main reasons.
First one is simply  for the better insight in to the asymptotic symmetry of the gravity theory.  The empty AdS is maximally symmetric solution, and studying the substructure of that solution due to the boundary conditions 
of the theory is a natural choice. 
The second reason is that, since we are considering conformal gravity whose equations of motion are not as restrictive as equations of motion of EG, the structure is richer than the structure in \cite{Henneaux:1985tv} and the structure in asymptotically flat space. 
Beside the global charges, 
we are interested in the local densities of the charges at the boundary and studyng entire boundary stress energy tensor. 
The third reason is that the conserved charges define ASA and play fundamental role for the AdS/CFT correspondence.

This article is structured as follows. In second and third section we review the general properties of CG and its asymptotic symmetries. We describe the procedure that leads to  ASA and classify its subalgebras.  In section four we analyse ASA of the known spherically symmetric solutions. In section five, we present highest subalgebras, while in section six we construct the global solutions from highest subalgeras, consider double holography  solution 
and global solution dependent on all the coordinates of the manifold, and in section seven we conclude. The appendices contain detailed classified subalgebras and analysis of the linear term.

\section{Conformal gravity}

Let us review the most important features of conformal gravity. The theory depends only on (Lorentz-) angles but not on the distances, i.e. it is invariant under local Weyl rescalings of the metric
\eq{
g_{\mu\nu} \to\tilde g_{\mu\nu} = e^{2\omega}\,g_{\mu\nu}\,, 
}{eq:CG1}
for Weyl factor $\omega$ that is allowed to depend on the coordinates on the manifold $\mathcal{M}$.
Conformal gravity bulk action 
\eq{
I_{\textrm{\tiny CG}} = \alpha_{\textrm{\tiny CG}} \int\extd^4x\sqrt{|g|} \, g_{\al\mu}g^{\beta\nu}g^{\ga\lambda}g^{\de\tau}\,C^\al{}_{\beta\ga\de} C^\mu{}_{\nu\lambda\tau}
}{eq:CG2}
consists of the square of the Weyl tensor $C^\al{}_{\beta\ga\de}$, and since $\omega$ from the square root of determinant is canceled with $\omega$ from the metric terms, the action is manifestly invariant under the Weyl rescalings.
The action yields equations of motion (EOM) fourth order in derivatives, which require vanishing of the Bach tensor 
\eq{
\big(\nabla^\delta\nabla_\ga + \frac12\,R^\delta{}_\ga\big)\, C^\ga{}_{\al\de\beta} = 0\,.
}{eq:CG3}

\section{Asymptotic symmetries in conformal gravity }

Following the notations of \cite{Grumiller:2013mxa}
we are interested to find the boundary conditions that preserve the gauge transformations 
\begin{equation} 
\delta g_{\mu\nu}=\left(e^{2\omega}-1\right)g_{\mu\nu}+\pounds_{\xi}g_{\mu\nu}. \label{ke}
\end{equation}
where the first term corresponds to Weyl transformations of the metric \ref{eq:CG1}
and the second term corresponds to small diffeomorphisms $x^{\mu}\rightarrow x^{\mu}+\xi^{\mu}$ generated by the vector field $\xi$.

We use the gauge
\begin{equation}
ds^2=\frac{1}{\rho^2}\left(-\sigma d\rho^2+\gamma_{ij}dx^idx^j \right) \label{metric}
\end{equation}
for $\sigma=\pm1$ for (AdS/dS) case and we expand the metric $\gamma_{ij}$ (metric at the boundary $\partial \mathcal{M}$), in the Fefferman-Graham (FG) expansion
\begin{equation}
\gamma_{ij}=\gamma_{ij}^{(0)}+\rho \gamma_{ij}^{(1)}+\frac{1}{2}\rho^2\gamma_{ij}^{(2)}+...\label{fge}
\end{equation}
We analogously expand the Weyl factor and the small diffeomorphisms
\begin{align}
\omega&=\omega^{(0)}+\rho \omega^{(1)}+\frac{1}{2}\omega^{(2)}\rho^2+... \\
\xi^{\mu}&=\xi^{\mu(0)}+\rho\xi^{\mu(1)}+\frac{1}{2}\xi^{\mu(2)}\rho^2+... .
\end{align}
The imposition of the conditions to the metric $\delta g_{\rho\rho}=0$ and the $\delta g_{\rho i}=0$ and keeping of form of the asymptotic expansion of the metric, leads to vanishing of the leading term in the expansion of the Weyl factor, $\omega^{(0)}=0$ and $\xi^{(0)}=0$. That gives
\begin{align}
\omega &=\rho \omega^{(1)} +\frac{1}{2}\rho^2\omega^{(2)}+... \\
\xi^{\rho}&=\rho\lambda -\rho^2\omega^{(1)}+... \\
\xi^i&=\xi^i_{(0)}+\frac{1}{2}\sigma \rho^2\mathcal{D}^i\lambda \label{xieq}
\end{align}
where $\lambda$, $\omega^{(n)}$ and $\xi^{i(0)}$ are arbitrary functions of the coordinates on the $\partial \mathcal{M}$ and $\mathcal{D}_i$ is covariant derivative on $\partial \mathcal{M}$ compatible with the metric $\gamma_{ij}^{(0)}$. 
The boundary conditions \cite{Grumiller:2013mxa}
\begin{align}
\delta\gamma_{ij}^{(0)}&=2\overline{\lambda}\gamma_{ij}^{(0)}\label{bcs1}\\  \delta\gamma_{ij}^{(1)}&=\overline{\lambda}\gamma_{ij}^{(1)}\label{bcs2}\end{align}  are defined by $\overline{\lambda}$, taking into an account with an automatic assumption that the expansion of the metric does not contain any logarithmic terms. 
These boundary conditions were defined in \cite{Irakleidou:2014vla}  and earlier studied in \cite{Grumiller:2013mxa}. Generally, the boundary conditions are chosen such that the variational principle is well defined and the response functions are finite. While searching for the right boundary conditons one has to keep in mind not to overrestrict the theory while obtaining the above properties. 
 In the expansion of the $\xi^i$ (\ref{xieq}) we must notice that there is no linear term, which means that the transformation of the $\gamma_{ij}^{(1)}$  will depend only on the $\xi_{(0)}^i$, $\lambda$ and $\omega^{(1)}$.
In the leading and the subleading order in $\rho$ respectively, the equation (\ref{ke}) will give equations 
\begin{align}
\delta \gamma_{ij}^{(0)}&=\mathcal{D}_i\xi_j^{(0)}+\mathcal{D}_j\xi^{(0)}_i-2\lambda\gamma_{ij}^{(0)}\label{eqnn1}\\ 
\delta\gamma_{ij}^{(1)}&=\pounds_{\xi^{k}_{(0)}}\gamma_{ij}^{(1)}+4\omega^{(1)}\gamma_{ij}^{(0)}-\lambda\gamma_{ij}^{(1)}.\label{eqnn2}
\end{align}
Taking the trace of (\ref{eqnn1}) and using (\ref{bcs1})
leads to 
\begin{equation}\overline{\lambda}=\frac{1}{3}\gamma_{il}^{(0)}\mathcal{D}^i\xi^{l}-\lambda,\end{equation}
and then the leading order conformal Killing equation (\ref{eqnn1}) becomes  \begin{equation}
\mathcal{D}_{i}\xi^{(0)}_j+\mathcal{D}_j\xi^{(0)}_i=\frac{2}{3}\gamma_{ij}^{(0)}\mathcal{D}_{k}\xi^{(0)k}\label{loke}.
\end{equation} 
Which one may recognise as a known Killing equation that defines ASA for EG, i.e. conformal algebra. Determining $\omega^{(1)}$ by taking the trace of (\ref{eqnn2}) and using the definition of $\lambda$ the subleading 
equation (\ref{eqnn2}) is
\begin{equation}
\pounds_{\xi^{(0)}}\gamma_{ij}^{(1)}=\frac{1}{3}\mathcal{D}_k\xi_{(0)}^k\gamma_{ij}^{(1)}.\label{nloke}
\end{equation}

In the leading order equation (\ref{loke}) we can choose the background metric $\gamma_{ij}^{(0)}$ and obtain the set of asymptotic Killing vectors (KVs) $\xi_i^{(0)}$. 
And the subleading order equation (\ref{nloke}) for the given subset of Killing vectors defines $\gam$ or we can define $\gam$ to obtain these Killing vectors.
 
 \subsection{Analysis of the leading order conformal Killing equation}
 
We start the analysis of the asymptotic symmetry algebra with the analysis of the equation (\ref{loke}). 
We search for the KVs $\xi^{(0)}$ by imposing $\gamma_{ij}^{(0)}$ to be
\begin{itemize}
\item
 the flat background 
 \begin{equation}
 \gamma_{ij}^{(0)}=\eta_{ij}=diag(-1,1,1)
 \end{equation}
 defined with the coordinates  $(t,x,y) $ on the boundary of the manifold $\partial\mathcal{M}$, on which we focus here,
 \item and the spherical $\mathbb{R}\times S^2$ background
 \begin{equation}
 \gamma_{ij}^{(0)}=\eta_{ij}=diag(-1,1,\sin(\theta^2))
 \end{equation}
with coordinates $(t,\theta,\phi)$, to which one can map the solution from the flat background.
\end{itemize}
For the flat background, KVs have been computed
in the standard literature on the conformal field theories \cite{DiFrancesco:1997nk}. 
In three dimensional Minkowski space these KVs are
\begin{align}
 \xi^{(0)} &= \partial_t, &
 \xi^{(1)} &= \partial_x, &
 \xi^{(2)} &= \partial_y  \label{translat}\\ 
\xi^{(3)} &= x\partial_t + t\partial_x &
 \xi^{(4)} &= y\partial_t + t\partial_y &
 \xi^{(5)} &= y\partial_x - x\partial_y
\end{align}
 \begin{align}
  \xi^{(6)} &= t\partial_t + x\partial_x + y\partial_y
 \end{align}
  \begin{align}
  \xi^{(7)} &= tx\partial_t + \frac{t^2+x^2-y^2}{2}\,\partial_x + xy\partial_y \\
  \xi^{(8)} &= ty\partial_t + xy\partial_x + \frac{t^2+y^2-x^2}{2}\,\partial_y \\
 \xi^{(9)} &= \frac{t^2+x^2+y^2}{2}\,\partial_t + tx\partial_x + ty\partial_y \label{origckvs}.
\end{align}
 three generators of translations (Ts: $\xi^{t}=(\xi^{(0)},\xi^{(1)}),\xi^{(2)})$),  which together with the generators of the Lorentz rotations  (L. rotations: $ L_{ij}=(x_i\partial_j-x_j\partial_i)$) for  ($\xi^{(3)},\xi^{(4)},\xi^{(5)}$), form Poincare algebra, dilatations (Ds:$\xi^{(6)}\equiv\xi^d$), and special conformal transformations (SCTs: $\xi^{sct}=(\xi^{(7)},\xi^{(8)},\xi^{(9)})$). They close into conformal algebra $so(3,2)$
\begin{align}
[\xi^d,\xi^t_j]&=-\xi^t_j && [\xi^d,\xi^{sct}_j]=\xi^{sct}_j\\
[\xi_l^t,L_{ij}]&=(\eta_{li}\xi^t_j-\eta_{lj}\xi^t_i) && [\xi_{l}^{sct},L_{ij}]=-(\eta_{li}\xi^{sct}_{j}-\eta_{lj}\xi^{sct}_i) \label{ca1}
\end{align}
\begin{align}
[\xi_i^{sct},\xi_j^t]&=-(\eta_{ij}\xi^d-L_{ij})\\
[L_{ij},L_{mj}]&=-L_{im}\label{ca2}
\end{align}
which we can verify explicitly. The vanishing of the linear term in the FG expansion (\ref{fge}) leads to the vanishing of the subleading Killing equation (\ref{nloke}) which means the asymptotic symmetry algebra is pure conformal algebra.
If the linear term in (\ref{fge}) does not vanish, one needs to analyse ASA using (\ref{nloke}), which can be performed in one of the two ways
\begin{enumerate}
\item First can be called "top-down" approach in which we bring full solution of CG to FG form for flat $\gamma_{ij}^{(0)}$. That defines $\gamma_{ij}^{(1)}$ that through (\ref{nloke}) determine the conserved KVs, i.e. corresponding subalgebra.
\item  \begin{enumerate} \item Impose restrictions on $\gamma_{ij}^{(1)}$ and determine the subalgebra from (\ref{nloke}),  \item impose subalgebra and determine $\gamma_{ij}^{(1)}$ .\end{enumerate}
\end{enumerate}
We can also define a linear combination of the KVs
\begin{align}
\xi^{lc}&=a_0\xi^{(0)}+a_1\xi^{(1)}+a_2\xi^{(2)}+a_3\xi^{(3)}+a_4\xi^{(4)}+a_5\xi^{(5)}+a_6\xi^{(6)}+a_7\xi^{(7)}\nonumber \\ &+a_8\xi^{(8)}+a_9\xi^{(9)} \label{lc},
\end{align}
and repeat the procedure under point 2. In the case 2.b. the imposed subalgebra can consist from 
\begin{enumerate}
\item the selected KVs, 
\item subalgebras classified according to the \cite{Patera:1976my}.
\end{enumerate}

 \section{Asymptotic symmetry algebras of the known global solutions}

Let us consider first the ASAs of the known solutions. For that we use the first "top-down" approach.
 That allows us to compute the corresponding response functions, energy, entropy and the momentum. 
  Most general spherically symmetric solution of CG, Mannheim--Kazanas--Riegert (MKR) solution \cite{Mannheim:1988dj} is defined by 
\eq{
\extd s^2 = -k(r)\extd t^2+\frac{\extd r^2}{k(r)}+r^2\,\extd\Om^2_{S^2}
}{eq:CG4}
for $d\Omega_{S^2}^2$ the line-element of the two sphere, and
\eq{
k(r)=\sqrt{1-12aM}-\frac{2M}{r}-\Lambda r^2+2ar.
}{eq:CG5}
 The response functions are $\tau^i{}_{j}=-8mp^i{}_j+8aa_Mdiag(1,-1,-1)^i{}_j$ for $p^i{}_j=diag(1,-\frac{1}{2},-\frac{1}{2})^i{}_j$ and $m=M$, while $P^i{}_j=8a_Mp^i{}_j$. Here, $a_M=\frac{1-\sqrt{1-12aM}}{6}$ and $\sigma=-1$ \cite{Grumiller:2013mxa}. 
This metric (33), we transform into FG expansion, so that $\gamma_{ij}^{(0)}$ is now $diag(-1,2,sin\theta^2)$ which corresponds to spherical background, and $\gamma_{ij}^{(1)}$ is
\begin{align}
\gam=\left(\begin{array}{ccc} 0 &0 &0 \\
0 & a & 0\\
0&0&a \sin^2\theta \end{array} \right)
\end{align}
which closes into the $\mathcal{R}\oplus o(3)$ algebra. $\partial_t$ is the only KV with non-vanishing charge $Q[\partial_t]=m-a a_M$ as an associated conserved charge for normalisation of the action $\alpha_{CG}=\frac{1}{64\pi}$. The entropy is $S=A_h/4$ with $A_h=4\pi r_h^2$ area of the horizon, which is $k(r_h)=0$.
\\

The second example of this approach is the rotating black hole. Its parametrisation consists of the Rindler acceleration $\mu$, rotation parameter $\tilde{a}$ and the vanishing mass parameter. That leads to zero PMR. The  Brown-York stress energy tensor gives for the conserved energy $E=-\tilde{a}\mu^2/[(1-\tilde{a})^2]$ and for the conserved angular momentum $J=E/\tilde{a}$, both of which are linear in the Rindler parameter $\mu$.

To write the solution in the form of the FG expansion, first we transform $\gamma_{ij}^{(0)}$ to $diag(-1,1,\sin^2\theta)$.
Applying  the same transformation to $\gamma_{ij}^{(1)}$ we obtain
\begin{align}
\gam=\left(
\begin{array}{ccc}
 \frac{4\mu}{2-\al^2+\al^2\cos(2\theta)} &0& \frac{4\al\mu\sin^2\theta}{2-\al^2+\al^2\cos(2\theta)} \\ 0 & 0 & 0\\ \frac{4\al\mu\sin^2\theta}{2-\al^2+\al^2\cos(2\theta)} &0 & \frac{4\al^2\mu\sin^4\theta}{2-\al^2+\al^2\cos(2\theta)}
\end{array}
\right)
\end{align}
which conserves $\partial_{\phi}$ and $\partial_t$ KVs of an Abelian $o(2)$ algebra.

Third example is the Schwarzschild BH \cite{Henneaux:1985tv}. Transforming it into FG form leads to vanishing $\gam$ which means that ASA is entire $o(3,2)$. The analysis of the conserved stress energy tensor agrees with the one in \cite{Maldacena:2011mk} and \cite{Deser:2002rt}.

Above global solutions of CG: MKR, rotating BH, and Schwarzschild BH, lead to ASAs that are 4-dimensional, 2-dimensional and full $o(3,2)$ respectively.
 In the following section we analyse the opposite direction, knowing the asymptotical solution, we consider which is the corresponding global solution. 
 



\section{Highest asymptotic symmetry algebras}
In this section we describe the highest dimensional subalgebras of CG  ASA. 
As we found from (17), ASA of CG is conformal algebra for the trivial $\gamma^{(1)}=0$ case. The classification of the highest dimensional subalgebras of CG ASA is classification of subalgebras with the most Killing vectors. 
 We impose the additional condition of tracelessness to $\gam$ since its trace is gauge. 
The dependency of the ASA on components of $\gam$ and coordinates on $\partial\mathcal{M}$ is given in the appendix: Dependency on coordinates.

\subsection{Five dimensional algebra}

In one dimension lower, $AdS_3/CFT_2$ correspondence, properties of two dimensional QFT \cite{Belavin:1984vu} are guided by Virasoro algebra whose investigation leads to learning about the $AdS_3$ and the correspondence itself. 
However, symmetries from two dimensions are not inherited in $CFT_3$, therefore it is of interest to study ASAs that define field theory in 3 dimensions.
In addition, the recent paper by Maldacena showed that imposing the Neumann boundary conditions to CG, gives EG solutions (which  in our notation corresponds to setting $\gamma^{(1)}$ to zero). To obtain the better insight into gauge/gravity correspondence, allowed algebras at the boundary, and possibly simpler way of finding general solutions, it is of interest to analyse the algebras dictated by the term $\gamma^{(1)}$. 
The highest realised algebra is 5-dimensional subalgebra of $o(3,2)$ $o(2)\ltimes o(1,1)$ (or $a_{5,4}=b_{5,6}$ in notation of \cite{Patera:1976my}). On the example we present, it is realised via constant $\gam$ which means Ts are automatically conserved. The additional KVs that form the algebra are two new linearly combined KVs
\begin{align}
\chi^{(1)}&=\xi^{(6)}-\frac{1}{2}\xi^{(3)} & \chi^{(2)}&=\xi^{(5)}-\xi^{(4)}
\end{align}
which conserve $\gam$ \begin{equation}
\ga^{(1)}_{ij}=\left(
\begin{array}{ccc}
 c & c & 0 \\
 c & c & 0 \\
 0 & 0 & 0 \\
\end{array}
\right)\label{five}.
\end{equation}
Permutation of the KVs that define L. rotations in new KVs leads to two new matrices

 \begin{align}
\ga^{(1)}_{ij}&=\left(
\begin{array}{ccc}
 0 & 0 & 0 \\
 0 & c & i c \\
 0 & i c & -c \\
 \end{array}
\right), & \ga^{(1)}_{ij}&=\left(
\begin{array}{ccc}
 -c & 0 & c \\
 0 & 0 & 0 \\
 c & 0 & -c \\ 
\end{array}
\right)\label{five2}\end{align}for the new KVs 

\begin{align}
\chi^{(1)}&=2i\xi^{(6)}+\xi^{(5)} & \chi^{(2)}&=\xi^{(3)}+i\xi^{(4)}
\end{align}
and 
\begin{align}
\chi^{(1)}&=\xi^{(5)}-\xi^{(3)} & \chi^{(2)}&=\xi^{(6)}+\frac{1}{2}\xi^{(4)}
\end{align}
respectively.
\noindent 
If we name the generators
\begin{equation}
P_0=-\xi^{(0)},P_1=\xi^{(1)},P_2=\xi^{(2)},F=\xi^{(6)},K_1=\xi^{(3)}, K_2=\xi^{(4)},L_3=\xi^{(5)},\label{simid}
\end{equation}
we find that they arrange in the subalgebra of the similitude algebra sim(2,1), which is one of the largest subalgebras of the $o(3,2)$. Namely,  
we obtain  "$a_{5,4}$" subalgebra $(F+\frac{1}{2}K_2,-K_1+L_3,P_0,P_1,P_2)$ that belongs to the 3 dimensional extended Poincare algebra 
\begin{align}
[\xi^d,\xi^t_j]&=-\xi^t_j\\
[\xi^t_l,L_{ij}]&=-(\eta_{li}\xi^t_j-\eta_{lj}\xi^t_i)\\
[L_{ij},L_{mj}]&=L_{im}.
\end{align}
Further we consider lower dimensional subalgebras of $o(3,2)$ that admit nontrivial $\gam$.

\subsection{Four dimensional algebras}

The second largest boundary ASAs are 4 dimensional as it is the case for the ASA of MKR which we have seen above.
Subalgebras of $o(3,2)$ that are also four dimensional  include $\gam$ matrices 
\begin{enumerate}
\item
\begin{align}
 \gamma_{ij}^{(1)}=\left(
\begin{array}{ccc}
 \frac{c}{2 x} & 0 & 0 \\
 0 & \frac{c}{x} & 0 \\
 0 & 0 & -\frac{c}{2 x} \\
\end{array}
\right) \label{poind}
\end{align}
that conserves 2 dimensional extended Poincare algebra which consists of 2Ts, L. rotation and D, $\xi^{(0)}, \xi^{(2)},\xi^{(4)},\xi^d$; 
\item 
\begin{align}
 \gamma_{ij}^{(1)}=\left(
\begin{array}{ccc}
 -\frac{c(t^2+2x^2)}{3(t^2-x^2)^{3/2}}& \frac{ctx}{(t^2-x^2)^{3/2}} & 0 \\
 \frac{ctx}{(t^2-x^2)^{3/2}} & -\frac{c(2t^2+x^2)}{3(t^2-x^2)^{3/2}} & 0 \\
 0 & 0 & \frac{c}{t\sqrt{t^2-x^^2}} \\
\end{array}
\right) \label{poind}
\end{align}
which conserves D, SCT: $\left(ty, xy,\frac{1}{2}\left(t^2-x^2+y^2\right)\right)$, L. rotation and T in the $y$ direction;
\item  $\gam$ that conserves 3 SCTs and rotations analogously to MKR case with Ts interchanged with SCTs is:
\begin{align}
\gamma^{(1)}_{11}&=-\frac{\left(t^4+4 \left(x^2+y^2\right) t^2+\left(x^2+y^2\right)^2\right) c}{\left(t^2-x^2-y^2\right)^3} \nonumber   \\
\gamma^{(1)}_{12}&=\frac{3 t x \left(t^2+x^2+y^2\right) c}{\left(t^2-x^2-y^2\right)^3}\nonumber \\
\gamma_{13}^{(1)}&=\frac{3 t y \left(t^2+x^2+y^2\right) c}{\left(t^2-x^2-y^2\right)^3} \nonumber \\
\gamma_{22}^{(1)}&=-\frac{\left(t^4+2 \left(5 x^2-y^2\right) t^2+\left(x^2+y^2\right)^2\right) c}{2 \left(t^2-x^2-y^2\right)^3} \nonumber\\
\gamma_{23}^{(1)}&=-\frac{6 t^2 \sqrt{x} y c}{\left(t^2-x^2-y^2\right)^3} \nonumber \\
\gamma_{33}^{(1)}&=-\frac{\left(t^4-2 \left(x^2-5 y^2\right) t^2+\left(x^2+y^2\right)^2\right) c}{2 \left(t^2-x^2-y^2\right)^3}\label{rsct}
\end{align}
where the remaining components of $\gam$ are zero. 
\end{enumerate}
\noindent The subalgebras above are formed by the KVs  of $o(3,2)$ from (21) to (26), while the further solutions with four dimensional subalgebras we list in the Appendix B and Appendix C, where we classify them according to the original KVs and according to the classification in \cite{Patera:1976my}. The latter classification requires identification of KVs using the linear combinations of KVs. 
 In the following chapter, we consider global solutions.

\section{Global solutions for asymptotic symmetry algebras}

In this chapter we build global solutions from particular asymptotic symmetry solutions. We first consider the solutions that belong to highest dimensional ASAs.


\subsection{Global solution from five dimensional algebra}

To build a global solution from the solution that belongs to five dimensional ASA from the section "Five dimensional algebra" we consider the ansatz metric
\begin{equation} 
ds^2=dr^2+(-1+cf(r))dx_i^2+2cf(r)dx_idx_j+(1+cf(r))dx_j^2+dx_k^2\label{geon}.
\end{equation}
It solves the Bach equation when $x_i=t,x_j=y,x_k=y$ and 
\begin{equation}
f(r)=c_1+c_2 r+c_3 r^2+c_4 r^3\label{sol}
\end{equation}
 while for $c_3$ and $c_4$ equal to zero, the metric is Ricci flat. 
 This is the case when the line element is (\ref{metric})$\times \rho^2$ in which $\gam$ is (\ref{five})$\cdot c_2$ and all higher terms in the FG expansion are set to zero. 
Due to conformal invariance we can multiply the line element (\ref{geon}) with $\frac{1}{r^2}$ to obtain the form (\ref{metric}) (taking $r\rightarrow \rho$) exactly and compute the response functions \cite{Grumiller:2013mxa} Brown--York stress energy tensor and partially massless response (PMR)
\begin{align}
\begin{array}{ccc}
\tau_{ij}=\left(\begin{array}{ccc}-c c_4& -c c_4 &0 \\ -c c_4 & -c c_4 &0 \\ 0& 0 & 0  \end{array}\right)
& \text{  and  } &
P_{ij}=\left(\begin{array}{ccc}  c c_3 & c c_3 & 0 \\ c c_3 & c c_3 &0 \\ 0 & 0 & 0   \end{array}\right)
\end{array}, \label{rspf}
\end{align}
respectively. 
The choice of $c_3$ and $c_4$ in the metric dictates whether response functions will vanish. The charges \cite{Grumiller:2013mxa} associated with the timelike KV (1,0,0) and the KV (0,1,0), are $-2cc_4$ per square unit of AdS radius, while the charge associated with the KV (0,0,1) vanishes. 
The charges associated with the KVs $(t-\frac{x}{2},-\frac{t}{2}+x,y)$ and $(-y,y,-t-x)$ are $cc_4(t+x) $ and zero, respectively.
The solution is not conformally flat, however Weyl squared and therefore the entropy are equal to zero, while polynomial invariants are finite. It defines a geon, or pp-wave solution, that have been studied in \cite{Wheeler:1955zz}, \cite{Melvin:1963qx}, \cite{Kaup:1968zz} and recently discussed in relation to the instability of AdS spacetime \cite{Horowitz:2014hja}.

\subsection{Global solution from four dimensional algebras}

Analogous geon solution can be obtained as well from the lower dimensional subalgebras for the constant $\gam$, and $\gam$ that depend on the coordinates of the boundary manifold. 
Let us consider first the constant case.

\subsubsection{Constant $\gam$}
Analogous ansatz as in five dimensional subalgebra of (\ref{geon})  for $x_i=t,x_j=y,x_i=x$ leads to $f(r)$
that solves Bach equation with $f(r)=c_1+c_2 r+c_3 r^2+c_4 r^3$. Setting $c_1=0$ and multiplying  (\ref{geon}) with $\frac{1}{r^2}$ leads to FG form of (\ref{geon}) from which we read out the matrix
\begin{equation}
\gam=\left(\begin{array}{ccc}c c_2&0&c c_2\\ 0&0&0\\c c_2&0&c c_2\\ \end{array}\right)
\end{equation}
It corresponds to $a_{4,10}^b$ group from \cite{Patera:1976my} and belongs to similitude $sim(2,1)$ algebra described in the appendix: "Patera et al. classification". Its response functions  read 
\begin{align}
\tau_{ij}&=\left(
\begin{array}{ccc}
 -c c_4 & 0 & -c c_4 \\
 0 & 0 & 0 \\
 -c c_4 & 0 & -c c_4 \\
\end{array}
\right), & 
P_{ij}&=\left(
\begin{array}{ccc}
 c c_3 & 0 & c c_3 \\
 0 & 0 & 0 \\
 c c_3 & 0 & c c_3 \\
\end{array}
\right).
\end{align}
The charges corresponding to KVs, defined with linear combinations of (\ref{translat})-(\ref{origckvs}), are $-2(-1+b)c(t+y)c_4$ for $F-b K_2$ KV, $-2cc_4$ for $P_0$, zero for $P_1$ and $2cc_4$ for $P_2$. b is arbitrary coefficient. 
The similar form of the solution (\ref{geon}) leads to the $\gam$ of the form \begin{equation}\left(\begin{array}{ccc}0&c c_2&0\\ c c_2&0&-c c_2\\0&-c c_2&0\\ \end{array}\right)\end{equation} for which the essential coefficients in f(r) that solve the Bach equation and make the solution non-trivial, are $c_1$ and $c$. Since the higher coefficients $c_3$ and $c_4$ are those responsible for the existence of the response functions, they as well as the charges in this example vanish.
\subsubsection{$\gam$ dependent on two coordinates}

Using the analogous procedure as above, global solution can be also obtained from four dimensional algebra that depends on the two coordinates of $\partial\mathcal{M}$ starting from the 
KVs $\chi_1= a(-P_0+P_2)$, $\chi_2=P_1$,  $\chi_3=L_3-K_1$, defined in (41), that conserve 
\begin{align}
\gam&=\left(\begin{array}{ccc} c_1\cdot b(t-y) & 0 &-c_1\cdot b(t-y) \\ 0&0 & 0 \\ -c_1\cdot b(t-y)&0& c_1\cdot b(t-y) \end{array}\right),
\label{globdepn}
\end{align}
 and solving the (\ref{nloke}) for one further KV.
\begin{enumerate}
\item Choice of  $\chi_4=F+K_2+\kappa(-P_0-P_2) $ leads to $b(t-y)=b_1\cdot e^{\frac{t-y}{2\epsilon}}$.
This algebra belongs to the $a_{4,12}$ algebra from \cite{Patera:1976my} which contains the KVs $\{F+K_2+\epsilon(P_0+P_2),-K_1+L_3,P_0-P_2, P_1\}$ (with $\epsilon=-\kappa$) and leads to 
$\gam$
\begin{align}
\gam=\left(
\begin{array}{ccc}
 \text{c}_1 e^{\frac{t-y}{2 \kappa }} b_1 & 0 & -\text{c}_1 e^{\frac{t-y}{2 \kappa }} b_1 \\
 0 & 0 & 0 \\
 -\text{c}_1 e^{\frac{t-y}{2 \kappa }} b_1 & 0 & \text{c}_1 e^{\frac{t-y}{2 \kappa }} b_1 \\
\end{array}
\right)
\end{align} and 
response functions 
\begin{align}
\tau_{ij}&=\left(
\begin{array}{ccc}
 -c_4 e^{\frac{t-y}{2 \kappa }} b_1 & 0 & c_4 e^{\frac{t-y}{2 \kappa }} b_1 \\
 0 & 0 & 0 \\
 c e^{\frac{t-y}{2 \kappa }} b_1 & 0 & -c_4 e^{\frac{t-y}{2 \kappa }} b_1 \\
\end{array}
\right), & P_{ij}&= \left(
\begin{array}{ccc}
 c_3 e^{\frac{t-y}{2 \kappa }} b_1 & 0 & -c_3 e^{\frac{t-y}{2 \kappa }} b_1 \\
 0 & 0 & 0 \\
 -c_3 e^{\frac{t-y}{2 \kappa }} b_1 & 0 & c_3 e^{\frac{t-y}{2 \kappa }} b_1 \\
\end{array}
\right).
\end{align}
\item For $\chi_4=F$, the function $b(t-y)$ takes the form $b(t-y)=\frac{b}{t-y}$
\item $\chi_4=-P_0-P_2$ makes $b\cdot(t-y)$ to be constant
\item  $ \chi_4=F-K_2$ leads to $b(t-y)=\frac{b_1}{(t-y)^{3/2}}$
\item $\chi_4=K_2$ provides  $b(t-y)=\frac{b_1}{(t-y)^2}$
\item $\chi_4=F+cK_2$ gives $b(t-y)=b_1\cdot(t-y)^{\frac{1-2c}{-1+c}}$
This implies that an arbitrary profile $b\cdot(t-y)$ breaks $a_{5,4}$ into $L_3-K_1,P_1,-P_0+P_2$. If we focus on studying the profile in (\ref{globdepn}) when $b(t-y)=(t-y)^{\beta}$ one obtains beside the known conserved KVs additionally:
\item  $\chi_4=F$ when $\beta=-1$
\item $\chi_4=K_2$ for $\beta=-2$
\item and $\chi_4=F +\frac{1+\beta}{2+\beta}K_2$.
\end{enumerate}


\subsection{Global solution from three dimensional subalgebra}

We give two examples for lower dimensional ASAs that lead to global solution. The first  one is dependent on the one coordinate on the boundary and the second one on two.
The ansatz for the global solution dependent on one coordinate on the boundary is the metric
\begin{equation} 
ds^2=dr^2+(-1+b(x)f(r))dt^2+dx^2+2b(x)f(r)dtdy+(1+b(x)f(r))dy^2\label{gldep1}.
\end{equation}
It satisfies Bach equation when the function $f(r)$ takes one of the following values
\begin{enumerate}
\item $f=c_1+c_2 r+c_3 r^2+c_4 r^3$  for $b=a_1+a_2 x$,
\item $f=c_1+c_2 r$ for $b=a_1+a_2 x +a_3 x^2+a_4 x^3$.
\end{enumerate}
Here we can read out the $\gamma_{ij}^{(i)}$ ($i=1,2,3$) matrices. In the first case the response functions vanish if we set $c_3=c_4=0$, while the second case has vanishing response functions and charges. 
Corresponding $\gam$ matrix defines ASA
 \begin{eqnarray}
\chi_1&=&(0,0,1),\\
\chi_2&=&(t-y,x,-t+y), \\
\chi_3&=&(1,0,0), \label{ricflat}
 \end{eqnarray}
 formed by two Ts and combination of the boost in $t-y$ plane and the dilatation or the  $a^c_{3,19},c\neq0,\pm1,-2$ for KVs $P_0,P_2,F-c K_2$ with $c=1$ in the notation of  \cite{Patera:1976my}. The response functions are
 \begin{align}\begin{array}{cc}\tau_{ij}=\left(
\begin{array}{ccc}
 -x c_4 & 0 & -x c_4 \\
 0 & 0 & 0 \\
 -x c_4 & 0 & -x c_4 \\
\end{array}
\right), & P_{ij}=\left(
\begin{array}{ccc}
 -x c_4 & 0 & -x c_4 \\
 0 & 0 & 0 \\
 -x c_4 & 0 & -x c_4 \\
\end{array}
\right)\end{array}
\end{align}
while the conserved charges of the KVs $\chi_3$, $\chi_1$, $\chi_2$ are $2xc_4$, $2xc_4$ and zero, respectively. Analogously as for the five dimensional subalgebra, one can obtain the Ricci flat solution, when
 $\gamma^{\FO}_{ij}=\left(\begin{array}{c c c} x & 0& x \\ 0& 0 &0 \\ x& 0 & x\end{array}\right).
$ 
\\
The following example is a global solution which depends on the two coordinates on the boundary, that can be brought to double holography-like form. 
The ansatz 
\begin{align} 
ds^2&=dr^2+\left[-1+ b(t+x)f(r)\right] dt^2+2b(t+x)f(r)dtdx \nonumber \\&+ \left[1+ b(t+x)f(r)\right]dx^2+dy^2 \label{dep2ex1} 
\end{align}
solves the Bach equation when $f(r)=c_1+c_2 r+c_3r^2+c_4 r^4$
and leads to $\gam$
\begin{align}
\gam&=\left(\begin{array}{ccc} c_1b(t+x) & c_1b(t+x) &0 \\ c_1b(t+x)& c_1b(t+x) & 0 \\ 0&0& 0 \end{array}\right).
\label{globdep2}
\end{align} The corresponding linearly combined KVs are 
\begin{align}
\begin{array}{ll}
\chi_1&=(-1,1,0) \\
\chi_2&=(0,0,1) \\
\chi_3&=(-y,y,-t-x)
\end{array}\end{align}
while the response functions
\begin{align} \tau_{ij}&=\left(
\begin{array}{ccc}
 -c_4 b(t+x) & -c_4 b(t+x) & 0 \\
 -c_4 b(t+x) & -c_4 b(t+x) & 0 \\
 0 & 0 & 0 \\
\end{array}
\right), &  P_{ij}&=\left(
\begin{array}{ccc}
 c_3 b(t+x) & c_3 b(t+x) & 0 \\
 c_3 b(t+x) & c_3 b(t+x) & 0 \\
 0 & 0 & 0 \\
\end{array}
\right) \end{align}
lead to all three charges to vanish. 
To bring the solution to the double holography-like form we define the function 
$b(t+x)=\frac{2}{t+x}$ which brings to $\gam$ that conserves one additional KV, $\xi^{(6)}$.
We transform the coordinates $t\rightarrow\xi+\tau$ and $x\rightarrow\chi-\tau$ to obtain
\begin{equation}
ds^2=\frac{4 r d\chi^2}{\chi }+\frac{4 r d\chi dy}{\chi }+dr^2-4 d\tau d\chi +dy^2
\end{equation}
which upon transformation $r\rightarrow\chi\eta$ reads
\begin{equation}
ds^2=4 \eta  d\chi^2+(\chi  d\eta+\eta  d\chi)^2-4 d\tau d\chi +4 \eta  d\chi dy+dy^2.
\end{equation} 
One can analogously find the solution that considers function $b(t-y)$.

\subsection{Global solution dependent on three coordinates on the manifold}

With an ansatz 
\begin{equation} 
ds^2=dr^2+\left[-1+ b(t+x+y)f(r)\right] dt^2+dx^2+2b(t+x+y)f(r)dtdy + dy^2.
\end{equation}we obtain the solution to Bach equations when $f(r)=c_1+c_2 r+ c_3 r^2+c_4 r^3$ and $b(t+x+y)=b_1+b_2\cdot(t+x+y)$. The response functions are of analogous form as above  \begin{align}
 \tau_{ij}&=\left(
\begin{array}{ccc}
 -b_1-(t+x+y) b_2 & 0 & -b_1-(t+x+y) b_2 \\
 0 & 0 & 0 \\
 -b_1-(t+x+y) b_2 & 0 & -b_1-(t+x+y) b_2 \\
\end{array}
\right),\\ P_{ij}&=\left(
\begin{array}{ccc}
 \left(b_1+(t+x+y) b_2\right) c_4 & 0 & \left(b_1+(t+x+y) b_2\right) c_4 \\
 0 & 0 & 0 \\
 \left(b_1+(t+x+y) b_2\right) c_4 & 0 & \left(b_1+(t+x+y) b_2\right) c_4 \\
\end{array}
\right)
 \end{align}
and they conserve
$\chi^{(1)}=(-1,1,0)$ with the corresponding charge  $-2c_3\cdot\left[c_1+c_2\cdot(t+x+y)\right]$,  and $\chi^{(2)}=(-1,0,1)$ whose charge vanishes. The KVs form Abelian $o(2)$ algebra.


\section{Conclusion}

We have considered ASA of the CG holography in four dimensions. In the leading order of the Killing equation expansion with respect to holographic coordinate $\rho$ the ASA is equal to ASA of EG, conformal algebra $o(3,2)$ \cite{Henneaux:1985tv}. When we take into consideration the subleading order of the Killing equation (\ref{nloke}), ASA can be classified according to subalgebras of $o(3,2)$.
This classification was done depending on whether the non-trivial $\gam$ is realised (exists) for certain subalgebra. 
The largest realised subalgebra is $o(2)\ltimes o(1,1)$ (or $a_{5,4}$ from \cite{Patera:1976my}) and it contains 5 KVs.

We have constructed global from the asymptotic solutions for the highest $o(2)\ltimes o(1,1)$  subalgebra and obtained geon or pp wave solution with non-vanishing charges, however vanishing entropy. 
The size of ASA or coordinates on which corresponding $\gam$ depends, do not impose restriction on the construction of these global solutions. We have constructed geon solutions also from four (studied as well in in \cite{Irakleidou:2014vla} and in \cite{Lovrekovic:2015jsa}), three and two dimensional ASA where in the latest case $\gam$ depend on all the coordinates on the manifold. Geons can be described as smooth horizonless geometries, and it was shown in \cite{Dias:2015rxy}  that they can be connected to an onset of the super radiance instability, which translates into connecting the nonlinear, weakly turbulent instability of AdS to super radiant instability of Kerr-AdS.

The ASAs of MKR and rotating black hole are four dimensional $R\times o(3)$ and two dimensional $o(2)$ algebras, respectively, and give non-vanishing charges and entropy considered in \cite{Grumiller:2013mxa}.  Analogously, the solutions 
with 5 and 4 KVs in ASA can be presented with one of the above $\gam$ matrices, while all solutions of CG can be classified according to the above. 
%
In lower dimensions, asymptotic symmetry algebras have been studied in \cite{Troessaert:2013fma} and for the non-AdS higher spin gauge theories in \cite{Riegler:2012fa}.

As a further research one can look for the global solution, where ASAs are used to model the final ansatz that is inserted into Bach equation, however depending on the ansatz, further restrictions can be needed.
One may as well further analyse the known solutions. 
That includes computation of the thermodynamcal quantities as charges, entropy and mass in the holography with or without $\gam$. This  classifies thermodynamical quantities. 
The concept of mass in gravity theory measures the wrapping of the space which has for the AdS space been considered in \cite{Brown:1992br,Balasubramanian:1999re} and for the dS space in \cite{Balasubramanian:2001nb}.

One can as well compute the higher point functions taking as a background not only flat space or $R\times S^2$ but include the linear term in the background metric. Two point functions would then be classified according to the above classification.  Taking into account the perturbed background, it would also be interesting to see whether the existence of the general linear term as a part of the background implies non-unitarity of the theory, or unitarity depends on the choice of $\gam$ term. 

Analgously to computation of the higher point functions with the background metric which includes the linear term, one could consider the partition function \cite{Irakleidou:2015exd} with these backgrounds and obtain  their classification.  


\section{Acknowledgements}
We would like to thank Daniel Grumiller and Robert McNees for the guidance during the project, as well as in the analysis of the five dimensional pp-wave solution, and Daniel to pointing out the appearance of the double-holography like solution. We would also like to thank Florian Preis for the useful discussions in the analysis of MKR solution particularly for indicating the existence of transformation to the global solution with zero mass which leads to vanishing $\gamma_{ij}^{(2)}$ in the FG expansion. We would also like to thank Daniel, Robert and Florian for useful discussions during this research. 
The project was funded by the  START project Y 435-N16, project I 952-N16 and P 26328-N27 by the Austrian Science Fund (FWF) and by the Forschungsstipendien 2015 by Technische Universitaet Wien.

\section{Appendix}
\section{Appendix A: Dependency on coordinates}
We consider the form of the $\gamma_{ij}^{(1)}$ matrix depending on its components, following with the 2. (a) and 2. (b) cases in subsection "Analysis of the leading order conformal Killing equation".

General form of the $\gamma_{ij}^{(1)}$ can depend on all the coordinates on $\partial{\mathcal{M}}$ which a priori does not satisfy (\ref{nloke}). 

\textbf{Translations.} First we consider translational KVs. They are conserved if
\begin{equation}
\partial_k\gamma_{ij}^{(1)}=0, \label{condt}
\end{equation}
for $i,j,k=(t,x,y)\in\partial{\mathcal{M}}$, which
implies that $\gam$ dependent on three co-ordinates cannot conserve any translational KV. 
For translational KV to be conserved $\gam$ needs to be constant in the corresponding coordinate. 
For example, $\gam$ that conserves $\partial_x$ has components $f_{ij}(y,t)$ for $i,j=x,y,t $ components. 
Analogously, to conserve the translations in two directions, $\gamma_{ij}^{(1)}$ must not depend on the coordinates in these directions.

\textbf{Lorentz rotations.} The conservation of the Lorentz rotational KVs exhibits similar behaviour as conservation of Ts. For the diagonal $\gam$ to conserve the L.rotation around the y,x,t directions respectively, i.e. two boosts and rotations, the form of the $\gam$ needs to be 
\begin{align}
  \left(\begin{array} {ccc}f_1(y) f_2\bigl(-t^2+x^2\bigr)& 0 & 0 \\ 0 & -f_1(y)f_2\bigl(-t^2+x^2\bigr) &0 \\ 0 & 0 & 2f_1(y)f_2\left(-t^2+x^2\right)  \end{array}\right)  & \text{ for } \kvth,\label{trots0}\\
    \left(\begin{array} {ccc} f_1(x)f_2\bigl(-t^2+y^2\bigr)& 0 & 0 \\ 0 & 2f_1(x)f_2\bigl(-t^2+y^2\bigr) &0 \\ 0 & 0 & -f_1(x)f_2\left(-t^2+y^2\right)  \end{array}\right)  & \text{ for } \kvfo,\label{trots1} \\
  \left(\begin{array} {ccc} 2f_1(t)f_2\bigl(x^2+y^2\bigr)& 0 & 0 \\ 0 & f_1(t)f_2\bigl(x^2+y^2\bigr) &0 \\ 0 & 0 & f_1(t)f_2\left(x^2+y^2\right)  \end{array}\right)  & \text{ for } \kvfi. \label{trots}
 \end{align}
Here, we see asymptotic behaviour of components (analogous behaviour is present for the off-diagonal components of $\gam$.)
When $\gam$ is allowed to depend only on two coordinates, the only conserved L. rotational KVs can be the one with $f_2$ dependent on those coordinates, while $f_1$ has to be constant. 
For $\gam$ dependent on one coordinate, the only conserved L. rotational KV is the one with $f_1$ dependent on the corresponding coordinate, while $f_2$ has to be constant. 
The ratio of the $\gamma_{11}^{(1)}:\gamma_{22}^{(1)}:\gamma_{33}^{(1)}$ needs to be of the form (\ref{trots0}), (\ref{trots1}) to (\ref{trots}). 

One can simultaneously conserve two or three L. rotations. Consider the most restrictive case for $\gam$ that conserves three L. rotations. 
 $\gam$ that conserves 3 L. rotations is
\begin{equation}
\gamma_{ij}^{(1)}=\left(
\begin{array}{ccc}
 c \left(2 t^2+x^2+y^2\right) & -3 c t x & -3 c t y \\
 -3 c t x & c \left(t^2+2 x^2-y^2\right) & 3 c x y \\
 -3 c t y & 3 c x y & c \left(t^2-x^2+2 y^2\right) \\
\end{array}
\right)\label{only3rot}
\end{equation}
for $c$ arbitrary parameter. 

\textbf{Dilatations. }
Most general form of $\gam$ that conserves dilatations contains components of the form
\begin{equation}\ga^{\FO}_{ij}=\frac{b_{ij}\left(\frac{x_{j}}{x_{i}},\frac{x_{k}}{x_{i}}\right)}{x_{i}}+\frac{c_{ij}\left(\frac{x_{i}}{x_{j}},\frac{x_{k}}{x_{j}}\right)}{x_{j}}+\frac{d_{ij}\left(\frac{x_{i}}{x_{k}},\frac{x_{j}}{x_{k}}\right)}{x_{k}}\label{dilat2},\end{equation} for $i,j,x=t,x,y$, $i\neq j\neq k$. 
From which we see that for $\gam$ dependent on two coordinates, e.g. $x_i $ and $x_j$  consist from $b_{ij}$ and $c_{ij}$ functions dependent on $\frac{x_j}{x_i}$ and $\frac{x_i}{x_j}$ respectively.
 $\gam$ dependent on one coordinate contains the components of the form $\frac{f_{ij}}{x_i}$ for $i$ the coordinate on which $\gam$ depends. 

\textbf{Special conformal transformations (SCTs).} 
Examples of $\gam$ matrices that conserve SCTs are:
 $\gam$ that conserves the KV $\left( \frac{1}{2}(t^2+x^2+y^2),tx,ty \right)$
is
\begin{align}
\gam=\left(\begin{array}{ccc}-\frac{f_1(x^2+y^2)f\left(\frac{y}{x},Log\left[-\frac{-t^2+x^2+y^2}{x}\right]\right)}{3txy} & 0& 0\\ 0 & \frac{f_1(x^2-2y^2)f\left(\frac{y}{x},Log\left[-\frac{-t^2+x^2+y^2}{x}\right]\right)}{3txy} & \frac{f_1f\left(\frac{y}{x},Log\left[-\frac{-t^2+x^2+y^2}{x}\right]\right)}{t} \\ 0 &\frac{f_1f\left(\frac{y}{x},Log\left[-\frac{-t^2+x^2+y^2}{x}\right]\right)}{t} & \frac{f_1(2x^2-y^2)f\left(\frac{y}{x},Log\left[-\frac{-t^2+x^2+y^2}{x}\right]\right)}{3txy} \end{array}\right)
\end{align}
for $f_1=e^{ArcTanh\frac{(t^2+x^2+y^2)}{(t^2-x^2-y^2)}}$. 
While the KV $\left(tx,xy,\frac{1}{2} \left( t^2-x^2+y^2 \right) \right)$ is conserved by the $\gam$ 
\begin{align}
\gam=\left( \begin{array}{ccc} -\frac{\left( t^2+2x^2\right) f\left( \frac{x}{t},-\frac{t^2+x^2+y^2}{t}\right)}{3t^2x} & \frac{f\left( \frac{x}{t},-\frac{t^2+x^2+y^2}{t} \right)}{3t^2x} & 0\\ \frac{f\left( \frac{x}{t},-\frac{t^2+x^2+y^2}{t} \right)}{3t^2x}& -\frac{\left( 2t^2+x^2 \right)f\left( \frac{x}{t},-\frac{-t^2+x^2+y^2}{t}\right)}{3t^2x} &0 \\ 0 & 0 & \frac{\left(t^2-x^2\right)f\left(\frac{x}{t},\frac{-t^2+x^2+y^2}{t}\right)}{3t^2x} \end{array} \right)
\end{align}
and the $\left( tx,\frac{1}{2}\left(t^2+x^2-y^2\right),xy \right)$ conservs the $\gam$ 
\begin{align}
\gam=\left( \begin{array}{ccc} -\frac{\left(t^2+2y^2\right)f\left(\frac{y}{t},\frac{-t^2+x^2+y^2}{t}\right)}{3t^2y} & 0 & \frac{f\left( \frac{y}{t},\frac{-t^2+x^2+y^2}{t} \right)}{t} \\ 0 & \frac{\left(t^2-y^2\right)f\left(\frac{y}{t},\frac{-t^2+x^2+y^2}{t}\right)}{t} &0  \\ \frac{f\left( \frac{y}{t},\frac{-t^2+x^2+y^2}{t} \right)}{t} & 0 & -\frac{\left(2t^2+y^2\right)f\left( \frac{y}{t}\frac{-t^2+x^2+y^2}{t}\right)}{3t^2y} \end{array} \right).
\end{align}
Here, we have for simplicity taken one of the components of $\gam$ to be zero. 


 \section{Appendix B: Classification according to  the generators of the conformal group}

Existance of the $\gam$ for certain set of KVs is determined by closing of their subalgebra  which we can show on an example. Assume we have one KV of translations, one of L. rotations and one of SCTs.  The Poisson bracket among SCT and T closes in $[\xi_i^{sct},\xi_j^t]=2(\eta_{ij}\xi^d-L_{ij})$ which for $i\neq j$ means $[\xi^{sct},\xi^{t}_j]=-2L_{ij}$. Therefore, L. rotation in the direction $i$ and $j$ also needs to be satisfied. 
Since $[\xi_i^{t},L_{ij}]=\xi_j^t$ and $[\xi_i^{sct},L_{ij}]=\xi_{j}^{sct}$, in order for algebra to close, we need $\xi_i^t$ and $\xi_j^{sct}$ that lead to a subalgebra with six KVs  $\xi^t_i,\xi^t_j,\xi^{sct}_i,\xi^{sct}_j,L_{ij},\xi^d$. 

In Table I and Table II, we present the realised subalgebras. The first row lists the KVs starting from Ts,  to L. rotations, Ds and SCTs, and their combinations, respectively. The second row writes whether the KVs close into existing subalgebra, while the third row presents the $\gam$ matrix of the asymptotic solution. The fourth row shows the number of KVs.

\begin{center}
\begin{table}
\caption{Classification according to original KVs}
\hspace{-0.65cm}\begin{tabular}{ |l | p{4.3 cm} | p{6.9 cm} | p{0.5cm}|}
\hline
Algebra & Name/existence(closing) & Realization &  \\
 \hline\hline
\hspace{0.18cm}1 T & $\exists$ & $\left(
\begin{array}{ccc}
 \text{$\gamma_{11}$}(x,y) & \text{$\gamma_{12}$}(x,y) & \text{$\gamma_{13}$}(x,y) \\
 \text{$\gamma_{12}$}(x,y) & \text{$\gamma_{22}$}(x,y) & \text{$\gamma_{23}$}(x,y) \\
 \text{$\gamma_{13}$}(x,y) & \text{$\gamma_{23}$}(x,y) & \text{$\gamma_{11}$}(x,y)-\text{$\gamma_{22}$}(x,y) \\
\end{array}
\right)$ & 1 \\
\hspace{0.18 cm}2 T & $\exists$ & $\left(
\begin{array}{ccc}
 \text{$\gamma_{11}$}(x) & \text{$\gamma_{12}$}(x) & \text{$\gamma_{13}$}(x) \\
 \text{$\gamma_{12}$}(x) & \text{$\gamma_{22}$}(x) & \text{$\gamma_{23}$}(x) \\
 \text{$\gamma_{13}$}(x) & \text{$\gamma_{23}$}(x) & \text{$\gamma_{11}$}(x)-\text{$\gamma_{22}$}(x) \\
\end{array}
\right)$ & 2\\
\hspace{0.18 cm}3 T & $\exists$ & $
 \left(
\begin{array}{ccc}
 c_1 & c_2 & c_3 \\
 c_2 & c_4 & c_5 \\
 c_3 & c_5& c_1-c_4 \\
\end{array} \right)$ & 3\\
\hspace{0.18 cm}1 T + 1 R &  $[\xi^t_{l},L_{ij}]=\eta_{li}\xi^t_{j}-\eta_{lj}\xi^t_{i}$,    
$\nexists$ for $l=i$ or $j$, $\exists$ for $l\neq i \neq j$   &  (example): eq. (\ref{trot}) & 2 \\
\hspace{0.18 cm}1T + D & $\exists$ & 
$\exists$: see explanation below eq. (\ref{dilat2})  &2\\
$\begin{array}{l}\text{1 T + 1 R}\\ \text{+ D}\end{array}$ & $\exists$ & $\left(
\begin{array}{ccc}
 \frac{c_1}{\sqrt{x^2+y^2}} & \frac{x c_2+y c_3}{x^2+y^2} & \frac{y c_2-x c_3}{x^2+y^2} \\
 \frac{x c_2+y c_3}{x^2+y^2} & \frac{c_1}{2 \sqrt{x^2+y^2}} & 0 \\
 \frac{y c_2-x c_3}{x^2+y^2} & 0 & \frac{c_1}{2 \sqrt{x^2+y^2}} \\
\end{array}
\right)$ There exist analogous matrices for the translations in the two remaining directions that depend, for the translation in the $l$ direction on the coordinates $i\neq l$ and $j\neq l$ &3\\
&& (example) & \\
$\begin{array}{l}\text{1T + D}\\ \text{+1 SCT}\end{array}$& $[\xi_i^{sct},\xi_j^t]=2(\eta_{ij}\xi^d-L_{ij})$, $\exists$ for $i=j$; sl(2)  &
$ \left(
\begin{array}{ccc}
 \frac{f\left(\frac{x}{t}\right)}{t} & -\frac{3 x f\left(\frac{x}{t}\right)}{t^2+2 x^2} & 0 \\
 -\frac{3 x f\left(\frac{x}{t}\right)}{t^2+2 x^2} & \frac{\left(2 t^2+x^2\right) f\left(\frac{x}{t}\right)}{t^3+2 x^2 t} & 0 \\
 0 & 0 & \frac{\left(x^2-t^2\right) f\left(\frac{x}{t}\right)}{t^3+2 x^2 t} \\
\end{array}
\right)$ example for  $i=j=y$
 & 3 \\
$\begin{array}{l}\text{1 T + 1 R} \\ \text{+ D+1 SCT}\end{array}$ &  $\exists$: $[\xi^t_i,\xi^{sct}_if]=2\xi^d$, $[\xi^t_l,L_{ij}]=0$, $[\xi^{sct}_l,L_{ij}]=0$ for $l\neq i$ and $l\neq j$;
sl(2)+u(1) & example for $\xi^t_y,\xi^{sct}_y,L_{xt},D$, see eq. (81) &4\\
\hspace{0.18 cm}2 T + 1 R & $\exists$: \text{2d Poincare} & $\exists$ see eq. (\ref{dep1}) &3\\
$\begin{array}{l}\text{2 T + 1 R}\\ \text{+ D}\end{array}$ & $\exists$: \text{2d Poincare +D} & $\exists$ see eq. (\ref{poind}) &3\\
\hline
\end{tabular}
\end{table}
\end{center}

\begin{center}
\begin{table}
\caption{Continuation of Table I.}
\hspace{-1.2cm}\begin{tabular}{ | l  | p{1 cm} | p{9.7 cm} | p{0.5cm}|}
\hline
\hspace{0.18 cm}2 T + D & $\exists$  & $\exists$ $\gamma_{ij}^{(1)}=\left(
\begin{array}{ccc}
 \frac{\text{c1}}{x} & \frac{\text{c2}}{x} & \frac{\text{c3}}{x} \\
 \frac{\text{c2}}{x} & \frac{\text{c4}}{x} & \frac{\text{c5}}{x} \\
 \frac{\text{c3}}{x} & \frac{\text{c5}}{x} & \frac{\text{c1}-\text{c4}}{x} \\
\end{array}
\right)$ &3\\ 
$\begin{array}{l}\text{2 T + 1 R} \\ \text{+ D + 2 SCT}\end{array}$ & $\exists$ & $\nexists$ &5\\
\hspace{0.18 cm}3 T + 1R & $\exists$: MKR & $\exists$ $\gamma_{ij}=\left(
\begin{array}{ccc}
 2 \text{c} & 0 & 0 \\
 0 & \text{c} & 0 \\
 0 & 0 & \text{c} \\
\end{array}
\right)$ &4\\
\hspace{0.18 cm}3 T + 3 R & $\exists$ & $\nexists$ &6\\

$\begin{array}{c}\text{3 T + 3 R} \\ \text{+ D}\end{array} $& $\exists$ & $\nexists$  &7\\
\hspace{0.18 cm}3 T + D & $\exists$ & $\nexists$ & 4\\

\hspace{0.18 cm}3 T + 3 R & $\exists$ & $\nexists$ &6\\

$\begin{array}{c}\text{3 T + 3 R}\\ \text{+ D}\end{array}$ & $\exists$ & $\nexists$ &7\\
\hspace{0.18 cm}3 T + D & $\exists$ & $\nexists$ & 4\\  

\hspace{0.18 cm}1 R & $\exists$ & $\exists$ see eq. (72) & 1 \\ 
\hspace{0.18 cm}3 R & $\exists$ & $\exists$ see eq. (\ref{only3rot})& 3 \\
\hspace{0.18 cm}1 R+D & $\exists$ & $\gamma_{ij}^{(1)}=\left(
\begin{array}{ccc}
 \frac{f\left(\frac{x^2+y^2}{2 t^2}\right)}{t} & \frac{a x+b y}{t^2} & \frac{a y-b x}{t^2} \\
 \frac{a x+b y}{t^2} & \frac{f\left(\frac{x^2+y^2}{2 t^2}\right)}{2 t} & 0 \\
 \frac{a y-b x}{t^2} & 0 & \frac{f\left(\frac{x^2+y^2}{2 t^2}\right)}{2 t} \\
\end{array}
\right)$ &2 \\ 
\hspace{0.18 cm}3 R+D & $\exists$  & $\exists$ & 4\\
\hspace{0.18 cm} 3 R+D & $\exists$  &$\exists$ see eq. (\ref{tn1})& 4 \\ 
\hspace{0.18 cm}1 R + 2 SCT &   $\exists$  & $\exists$ see eq. (\ref{eqq}) & 3\\ 
\hspace{0.18 cm}1 R + 3 SCT & $\exists$   & $\exists$ see eq. (\ref{rsct})  & 4\\ 
\hspace{0.18 cm}3R+3SCT & $\exists$   & $\nexists$ & 6 \\ 
\hspace{0.18 cm}1 R+D+2 SCT & $\exists$ & $\exists$ see eq. (\ref{t3}) & 4\\
$\begin{array}{c}\text{1 R+D}\\ \text{+3 SCT}\end{array}$ & $\exists$ & $\nexists$  & 5\\ 
$\begin{array}{c}\text{3 R+D}\\\text{+3 SCT} \end{array}$& $\exists$ & $\nexists$ & 7\\
\hspace{0.18 cm}1 SCT & $\exists$ &$\exists$ see eq. (\ref{t4}) & 1\\
\hspace{0.18 cm}2 SCT & $\exists$ & $\exists$ see eq. (\ref{t5}) & 2\\
\hspace{0.18 cm}3 SCT &$\exists$ &  & 3\\ 
\hspace{0.18 cm}1 SCT+D & $\exists$ & $\exists$ see eq. (\ref{t6}) & 1\\
\hspace{0.18 cm}2 SCT+D & $\exists$ &$\exists$ see eq. (\ref{t7}) & 2\\
\hspace{0.18 cm}3 SCT+D &$\exists$ & & 3\\ 
 \hline
\end{tabular}
\end{table}
\end{center}

 From the above table $\gam$ that conserves 1 T and 1 R, for T in $t$ direction and rotation in $x,y$ plane is \begin{align}
\gam=\left(
\begin{array}{ccc}
 f\left(\frac{1}{2} \left(x^2+y^2\right)\right) & a x+b y & a y-b x \\
 a x+b y & \frac{1}{2} f\left(\frac{1}{2} \left(x^2+y^2\right)\right) & 0 \\
 a y-b x & 0 & \frac{1}{2} f\left(\frac{1}{2} \left(x^2+y^2\right)\right) \\
\end{array}
\right)\label{trot},
\end{align}
while 
  $\gamma_{ij}^{(1)}$ that conserves $y$ translation, rotation around $y$ axis, D and SCT in $y$ direction is 
\begin{equation}
\gamma_{ij}^{(1)}=\left(
\begin{array}{ccc}
 -\frac{\left(t^2+2 x^2\right) c}{3 ((t-x) (t+x))^{3/2}} & \frac{t x c}{((t-x) (t+x))^{3/2}} & 0 \\
 \frac{t x c}{((t-x) (t+x))^{3/2}} & -\frac{\left(2 t^2+x^2\right) c}{3 ((t-x) (t+x))^{3/2}} & 0 \\
 0 & 0 & \frac{c}{3 \sqrt{(t-x) (t+x)}} \\
\end{array}
\right).\label{scty}
\end{equation}

$\gam$ that realises two Ts in the $x$ and $y$ directions and one L. R around the $t$ direction 
is 
\begin{align}
\gam=\left(\begin{array} {ccc} 2f(t) & 0 & 0 \\ 0 & f(t) &0 \\ 0 & 0 & f(t)  \end{array}\right)\label{dep1}
\end{align}

and  $\gam$ that realises 3 R and D is
 \begin{align}
\gamma_{11}^{(1)}&=  \frac{\left(2 t^2+x^2+y^2\right) c}{2 t^3 \left(-\frac{-t^2+x^2+y^2}{t^2}\right)^{3/2}} & &
\gamma_{12}^{(1)}=\frac{3 x c}{2 t^2 \left(-\frac{-t^2+x^2+y^2}{t^2}\right)^{3/2}}\nonumber \\
\gamma_{13}^{(1)}&= -\frac{3 y c}{2 t^2 \left(-\frac{-t^2+x^2+y^2}{t^2}\right)^{3/2}} &&
\gamma_{22}^{(1)}=\frac{\left(t^2+2 x^2-y^2\right) \sqrt{-\left(-t^2+x^2+y^2\right)} c}{2 \left(-t^2+x^2+y^2\right)^2}\nonumber \\
\gamma_{23}^{(1)}&=\frac{3 x y c}{2 t^3 \left(-\frac{-t^2+x^2+y^2}{t^2}\right)^{3/2}} &&
\gamma_{33}^{(1)}=\frac{\sqrt{-\left(-t^2+x^2+y^2\right)} \left(t^2-x^2+2 y^2\right) c}{2 \left(-t^2+x^2+y^2\right)^2}\label{tn1}
 \end{align}

\noindent The subalgebra with 1 R and 2 SCTs (e.g. rotation and SCTs in $x$ and $y$ direction) defines $\gam$
 \begin{align} 
\gamma_{11}^{(1)}&= \frac{\left(t^4+4 \left(x^2+y^2\right) t^2+\left(x^2+y^2\right)^2\right) f\left(\frac{-t^2+x^2+y^2}{t}\right)}{12 t^3} \nonumber \\ \nonumber
 \gamma_{12}^{(1)}&= -\frac{x \left(t^2+x^2+y^2\right) f\left(\frac{-t^2+x^2+y^2}{t}\right)}{4 t^2} \\ \nonumber
\gamma_{13}^{(1)}&= -\frac{y \left(t^2+x^2+y^2\right) f\left(\frac{-t^2+x^2+y^2}{t}\right)}{4 t^2}  \\ \nonumber
 \gamma_{22}^{(1)}&=\frac{\left(t^4+2 \left(5 x^2-y^2\right) t^2+\left(x^2+y^2\right)^2\right) f\left(\frac{-t^2+x^2+y^2}{t}\right)}{24 t^3}  \\ 
 \gamma_{23}^{(1)}&= -\frac{y \left(t^2+x^2+y^2\right) f\left(\frac{-t^2+x^2+y^2}{t}\right)}{4 t^2} \label{eqq}.
 \end{align}

 1 R, 2 SCTs and D are realised in the $\gam$
 
 \begin{align}
 \gamma_{12}^{(1)}& = -\frac{x \left(t^2+x^2+y^2\right) c}{4 \left(-t^2+x^2+y^2\right)^2} &&
 \gamma_{11}^{(1)}=\frac{\left(t^4+4 \left(x^2+y^2\right) t^2+\left(x^2+y^2\right)^2\right) c}{12 t \left(-t^2+x^2+y^2\right)^2} \nonumber \\
\gamma_{13}^{(1)}&= -\frac{y \left(t^2+x^2+y^2\right) c}{4 \left(-t^2+x^2+y^2\right)^2}  &&
\gamma_{22}^{(1)}= \frac{\left(t^4+2 \left(5 x^2-y^2\right) t^2+\left(x^2+y^2\right)^2\right) c}{24 t \left(-t^2+x^2+y^2\right)^2}  \nonumber \\
\gamma_{23}^{(1)}&=\frac{t x y c}{2 \left(-t^2+x^2+y^2\right)^2}, && \label{t3}
 \end{align}
\noindent here we solve (\ref{loke}) with $\xi^{(0)i}=D^{i}$ and $\gam$ (\ref{eqq}) (that has function $f$) for $f$. 
\noindent Possible $\gam$ that realises 1 SCT (in $y$ direction) is given with 
\begin{align}
\gamma_{11}^{(1)}&= -\frac{\left(t^2+2 x^2\right) f\left(\frac{x}{t},\frac{-t^2+x^2+y^2}{t}\right)}{3 t^2 x} &&\gamma_{12}^{(1)}=\frac{f\left(\frac{x}{t},\frac{-t^2+x^2+y^2}{t}\right)}{t} \nonumber \end{align} \begin{align}
\gamma_{22}^{(1)}&= -\frac{\left(2 t^2+x^2\right) f\left(\frac{x}{t},\frac{-t^2+x^2+y^2}{t}\right)}{3 t^2 x} \nonumber \\ \gamma_{33}^{(1)}&=\frac{(t-x) (t+x) f\left(\frac{x}{t},\frac{-t^2+x^2+y^2}{t}\right)}{3 t^2 x} , \label{t4}
\end{align}
$\gamma_{13}^{(1)}=\gamma_{23}^{(1)}=0$,
Due to $f\left(\frac{x}{t},\frac{-t^2+x^2+y^2}{t}\right)$ we can insert (\ref{t4}) in (\ref{nloke}) and find the further KVs.
(To avoid clutter we have written $\gam$  (\ref{t4}) that is not of the most general form, the most general form of the $\gam$ is given in \cite{thesis}.)
 $\gam$ that conserves SCT in $x$ direction is of the similar form as (\ref{t4})
\begin{align}
  \gamma_{11}^{(1)}&=   -\frac{\left(t^2+2 y^2\right) f\left(\frac{y}{t},\frac{-t^2+x^2+y^2}{t}\right)}{3 t^2 y}&&   \gamma_{13}^{(1)}=\frac{f\left(\frac{y}{t},\frac{-t^2+x^2+y^2}{t}\right)}{t} \nonumber \\  
\gamma_{22}^{(1)}&=  \frac{(t-y) (t+y) f\left(\frac{y}{t},\frac{-t^2+x^2+y^2}{t}\right)}{3 t^2 y} &&   \gamma_{33}^{(1)}=-\frac{\left(2 t^2+y^2\right) f\left(\frac{y}{t},\frac{-t^2+x^2+y^2}{t}\right)}{3 t^2 y}, \label{sctx}
\end{align}for $\gamma_{12}^{(1)}$ and $\gamma_{23}^{(1)}$ equal to zero. 
  The $\gam$ that conserves SCT in $t$ direction, computed using analogous simplifications as $\gam$ for SCT in $x$ and $y$ direction has the form 
 \begin{align}
 \gamma_{11}^{(1)}&=-\frac{e^{\tanh ^{-1}\left(\frac{t^2+x^2+y^2}{t^2-x^2-y^2}\right)} \left(x^2+y^2\right) f\left(\frac{y}{x},\log \left(-\frac{-t^2+x^2+y^2}{x}\right)\right)}{3 t x y}  
\nonumber \\
 \gamma_{22}^{(1)}&=\frac{e^{\tanh ^{-1}\left(\frac{t^2+x^2+y^2}{t^2-x^2-y^2}\right)} \left(x^2-2 y^2\right) f\left(\frac{y}{x},\log \left(-\frac{-t^2+x^2+y^2}{x}\right)\right)}{3 t x y}
 \nonumber \\
 \gamma_{23}&=\frac{e^{\tanh ^{-1}\left(\frac{t^2+x^2+y^2}{t^2-x^2-y^2}\right)} f\left(\frac{y}{x},\log \left(-\frac{-t^2+x^2+y^2}{x}\right)\right)}{\sqrt{t^2}}\label{scty} 
 \end{align}
\noindent $
 \gamma_{12}^{(1)}=0$, $\gamma_{13}^{(1)}=0$.
 Comparison of (\ref{t4}), (\ref{sctx}) and (\ref{scty}) acknowledges Minkowski background metric. 
$\gam$ that realises 2 SCTs (SCT in $y$ and $t$ direction) is
\begin{align}
 \gamma_{12}^{(1)}&=-\frac{(t+y) \left(x^2+(t+y)^2\right)}{2 x^2}&& \gamma_{11}^{(1)}=\frac{\left(x^2+(t+y)^2\right)^2}{4 x^3} \nonumber \\
\gamma_{13}^{(1)}&= -\frac{(t-x+y) (t+x+y) \left(x^2+(t+y)^2\right)}{4 x^3} &&\gamma_{22}^{(1)}=\frac{(t+y)^2}{x} \nonumber \\
\gamma_{23}^{(1)}&=\frac{(t+y) (t-x+y) (t+x+y)}{2 x^2} && \nonumber \\ \gamma_{33}^{(1)}&=\frac{(t-x+y)^2 (t+x+y)^2}{4 x^3} && \label{t5}
\end{align}
  while $\gam$ for 1 SCT and one D (SCT in $y$ direction) is
 \begin{align}
 \gamma_{11}^{(1)}&= \frac{t^2 y \left(t^2+x^2+y^2\right) f\left(\frac{x}{t}\right)}{2 x^2 \left(-t^2+x^2+y^2\right)^2} && \gamma_{12}^{(1)}=-\frac{t y \left(3 t^2+x^2+y^2\right) f\left(\frac{x}{t}\right)}{4 x \left(-t^2+x^2+y^2\right)^2}\nonumber \\
 \gamma_{13}^{(1)}&=-\frac{t \left(t^4+6 y^2 t^2-x^4+y^4\right) f\left(\frac{x}{t}\right)}{8 x^2 \left(-t^2+x^2+y^2\right)^2} && \gamma_{22}^{(1)}= \frac{t^2 y f\left(\frac{x}{t}\right)}{\left(-t^2+x^2+y^2\right)^2}\nonumber \\
 \gamma_{23}^{(1)}&=\frac{t^2 \left(t^2-x^2+3 y^2\right) f\left(\frac{x}{t}\right)}{4 x \left(-t^2+x^2+y^2\right)^2} && \gamma_{33}^{(1)}=\frac{t^2 y \left(t^2-x^2+y^2\right) f\left(\frac{x}{t}\right)}{2 x^2 \left(-t^2+x^2+y^2\right)^2} \label{t6}
 \end{align}
  and $\gam$ for 2 SCTs in $x$ and $y$ directions and D reads
 \begin{equation}
 \gam=\left(
\begin{array}{ccc}
 \frac{y \left(t^2+x^2+y^2\right) c}{2 \left(-t^2+x^2+y^2\right)^2} & -\frac{x y \left(3 t^2+x^2+y^2\right) c}{4 t \left(-t^2+x^2+y^2\right)^2} & -\frac{\left(t^4+6 y^2 t^2-x^4+y^4\right) c}{8 t \left(-t^2+x^2+y^2\right)^2}  \\
 -\frac{x y \left(3 t^2+x^2+y^2\right) c}{4 t \left(-t^2+x^2+y^2\right)^2} & \frac{x^2 y c}{\left(-t^2+x^2+y^2\right)^2} & \frac{x \left(t^2-x^2+3 y^2\right) c}{4 \left(-t^2+x^2+y^2\right)^2}  \\
 -\frac{\left(t^4+6 y^2 t^2-x^4+y^4\right) c}{8 t \left(-t^2+x^2+y^2\right)^2} & \frac{x \left(t^2-x^2+3 y^2\right) c}{4 \left(-t^2+x^2+y^2\right)^2} & \frac{y \left(t^2-x^2+y^2\right) c}{2 \left(-t^2+x^2+y^2\right)^2} 
\end{array}
\right).\label{t7}
 \end{equation}
 The largest realised subalgebra consisted of original KVs of $o(3,2)$ is four dimensional.  
That allows us to infer  $\gam$ for each of the KVs, and to anticipate form of $\gam$ for subalgebra of $o(3,2)$ classified according to Patera et al. Classification \cite{Patera:1976my}. 

\section{Appendix C: Patera et al. classification}
Here, we classify subalgebras of $o(3,2)$ for which (\ref{nloke}) has a $\gam$ solution. Algebraically, subalgebras of $o(3,2)$  have been classified in Patera et al. \cite{Patera:1976my} (we also follow their notation for groups). They are:
\begin{enumerate}
\item sim(2,1)  -- 7 dimensional similitude algebra, which realises 5 dimensional subalgebra
\item  opt(2,1) -- 7 dimensional optical algebra, which realises 5 dimensional subalgebra
\item $o(3)\oplus o(2)$ -- 4 dimensional maximal compact algebra 
\item $o(2)\oplus o(2,1)$ -- 4 dimensional algebra 
\item $o(2,2)$ -- 6 dimensional algebra 
\item $o(3,1)$ -- 6 dimensional Lorentz algebra which does not contain subalgbra with 5 generators.
\item $o(2,1)$ -- 3 dimensional irreducible algebra
\end{enumerate}
 To identify the generators of the subalgebras we linearly combine original generators of $o(3,2)$ (\ref{ca1},\ref{ca2}).
We consider which subalgebras have $\gam$ 
and focus on $\gam$ for the subalgebras with 7,6,5 and 4 generators.

\subsection{sim(2,1)}

Sim(2,1) is an algebra that consists of the seven generators that can be identified with (\ref{simid}), which is not unique identification of the generators. The realised subalgebras are classified in Table III and Table IV. The first column denotes the name of the subalgebra, the second and the third denote the name and generators as in \cite{Patera:1976my}, while the fourth column defines $\gam$. The definition of subalgebras according to Patera includes two subscripts. The first one is the dimension of the subalgebra and the second one enumerates the subalgebras of the same dimension. In each of the subalgebras,  first are listed the decomposable then undecomposable ones.  The parameter in the superscript denotes that algebra depends on a parameter, while the range of the parameter is denoted as $b\geq0,\neq1$ for $a_{4,10}^b$. When one range is written, it is equal under $O(3,2)$ and the identity component of the corresponding maximal subgroup (here $sim$(2,1)). For the range which is larger under the maximal subgroup than under $O$(3,2), the larger range is written with the square brackets, e.g. for $a_{4,8}^{\epsilon}$ it is written $\epsilon=1[\epsilon=\pm1]$ which means $a_{4,8}^{-1}$, conjugate to $a_{4,8}^1$ under $O$(3,2). Further details can be found in \cite{Patera:1976my}. 

\begin{center}
\begin{table}[ht!]
\caption{Classification according to \cite{Patera:1976my}, similitude algebra}
\hspace{-0.5cm}\begin{tabular}{ | l  | p{2.5 cm} | p{4,0 cm} | p{6.0cm} |}
 \hline
  \multicolumn{4}{|c|}{Realized subalgebras} \\
  \hline
$\begin{array}{c}\text{ Name/ }\\ \text{ commutators}\end{array}$&Patera name&generators  & Realisation  \\ \hline\hline
&$a_{5,4}^a$ & $\begin{array}{l}F+\frac{1}{2}K_2,-K_1+L_3,\\P_0,P_1,P_2\end{array}$  & see eq. (\ref{five2})
\\
&$a\neq0,\pm1$&$a=\frac{1}{2}$&\\
$\mathcal{R}\oplus o(3)$ &$a_{4,1}=b_{4,6}$ & $P_1\oplus\left\{K_2,P_0,P_2\right\}$  & $ \left(\begin{array}{ccc} \frac{c}{2} & 0 & 0 \\ 0 & c &0 \\ 0 & 0 & -\frac{c}{2}  \end{array}\right) $ \\

 &$a_{4,2}$ & $\begin{array}{l}P_0-P_2\oplus\\ \left\{F-K_2;P_0+P_2,P_1\right\}\end{array}$  &  $  \left(\begin{array}{ccc} -c & 0 & 0 \\ 0 & c &0 \\ 0 & 0 & -2c   \end{array}\right)$\\

$\begin{array}{c}
MKR \\  \mathcal{R}\oplus o(3)\end{array}$ &$a_{4,3}$ & $P_0\oplus\left\{L_3,P_1,P_2\right\}$  & $  \left(\begin{array}{ccc} 2c & 0 & 0 \\ 0 & c &0 \\ 0 & 0 & c   \end{array}\right) $ \\

 &$a_{4,4}$ & $F\oplus\left\{K_1,K_2,L_3\right\}$ & see eq. (\ref{tn1})\\

 &$a_{4,5}$ & $\begin{array}{l}F\{K_2;P_0-P_2\}\oplus \\ \left\{F-K_2,P_1\right\}\end{array}$ &$ \left(
\begin{array}{ccc}
 0 & \frac{c}{t-y} & 0 \\
 \frac{c}{t-y} & 0 & \frac{c}{y-t} \\
 0 & \frac{c}{y-t} & 0 \\
\end{array}
\right)$\\
&$a_{4,6}=b_{4,9}$ & 
$ \begin{array}{c} \left\{ F+K_2,P_0-P_2 \right\}\oplus  \\ \left\{F-K_2,P_0+P_2\right\} \end{array} $
 &
$\left(
\begin{array}{ccc}
 \frac{c}{x} & 0 & 0 \\
 0 & \frac{2 c}{x} & 0 \\
 0 & 0 & -\frac{c}{x} \\
\end{array}
\right)$,and (\ref{poind}) \\

&$a_{4,7}$&$ \begin{array}{c}  L_3-K_1,P_0+P_2 ;  \\ P_0-P_2,P_1 \end{array} $& leads to 5 KV subalgebra for constant components of $\gam$\\
\hline 
 \end{tabular}
 \label{tablesim}
 \end{table}
\end{center}

\begin{center}
\begin{table}[ht!]
\caption{Continuation of Table III.}
\hspace{-0.8cm}\begin{tabular}{ | l  | p{2.8 cm} | p{4.4 cm} | p{5.2cm} |}
\hline
&$\begin{array}{c}a_{4,10}^b=b_{4,13}\\ b>0,\neq1 \end{array}$ & $\left\{ F-bK_2,P_0,P_1,P_2 \right\}$  &  $\left(
\begin{array}{ccc}
 \text{c} & 0 & \text{c} \\
 0 & 0 & 0 \\
 \text{c} & 0 & \text{c} \\
\end{array}
\right), \left(
\begin{array}{ccc}
 0 & \text{c} & 0 \\
 \text{c} & 0 & -\text{c} \\
 0 & -\text{c} & 0 \\
\end{array}
\right) $, $\left(
\begin{array}{ccc}
 0 & \text{c} & 0 \\
 \text{c} & 0 & \text{c} \\
 0 & \text{c} & 0 \\
\end{array}
\right) $, $\begin{array}{l}\text{$\gam$ that leads to 5 KV}\\ \text{subalgebra}\end{array}$\\
 &$\begin{array}{c}a_{4,11}^b=b_{4,13}\\ b>0,[b\neq0] \end{array}$ & $\left\{ F+bL_3,P_0,P_1,P_2 \right\}$  &  $\left(
\begin{array}{ccc}
 0 & \text{c} & i \text{c} \\
 \text{c} & 0 & 0 \\
 i \text{c} & 0 & 0 \\
\end{array}
\right),\left(
\begin{array}{ccc}
 0 & \text{c} & -i \text{c} \\
 \text{c} & 0 & 0 \\
 -i \text{c} & 0 & 0 \\
\end{array}
\right),$ $\left(
\begin{array}{ccc}
 0 & 0 & 0 \\
 0 & i \text{c} & \text{c} \\
 0 & \text{c} & -i \text{c} \\
\end{array}
\right) $,  $\begin{array}{l}\text{$\gam$ that leads}\\ \text{to 5KV subalgebra}\end{array}$\\
 & $\begin{array} {c}a_{4,12}^{\epsilon}=b_{4,14}\\
 \epsilon=1*[\epsilon\pm1]
 \end{array}$ & 
$ \begin{array}{c}
 \big\{ F+K_2+\epsilon(P_0+P_2), \\
 -K_1+L_3,P_0-P_2,P_1 \big\}
 \end{array} $
&  $\left(
\begin{array}{ccc}
 ce^{\frac{y-t}{4 e}}  & 0 & -ce^{\frac{y-t}{4 e}} \\
 0 & 0 & 0 \\
 -ce^{\frac{y-t}{4 e}}  & 0 & ce^{\frac{y-t}{4 e}}  \\
\end{array}
\right) $\\
 &$a_{4,13}=b_{4,15}$ & $\begin{array}{l}\big\{ F-K_2,P_0-P_2,\\-K_1+L_3,P_1 \big\}\end{array}$   & $\left(
\begin{array}{ccc}
 \frac{c}{(y-t)^{3/2}} & 0 & -\frac{c}{(y-t)^{3/2}} \\
 0 & 0 & 0 \\
 -\frac{c}{(y-t)^{3/2}} & 0 & \frac{c}{(y-t)^{3/2}} \\
\end{array}
\right) $ \\

&$\tilde{a}_{4,14}$ & $\begin{array}{l}\big\{ F,-K_1+L_3,\\P_1,P_0-P_2 \big\}\end{array}$   & $\left(
\begin{array}{ccc}
 -\frac{c}{y-t} & 0 & \frac{c}{y-t} \\
 0 & 0 & 0 \\
 \frac{c}{y-t} & 0 & -\frac{c}{y-t} \\
\end{array}
\right)$ \\

 &$a_{4,15}=b_{4,17}$ & $\begin{array}{l}\big\{ F+bK_2,-K_1+L_3,\\P_0-P_2,P_1 \big\}\end{array}$ & 
$\begin{array}{l} \gamma_{11}^{(1)}=c\cdot(y-t)^{\frac{1-2 b}{b-1}},\\ \gamma_{13}^{(1)}=-c\cdot(y-t)^{\frac{1-2 b}{b-1}},\\ \gamma_{33}^{(1)}=c\cdot(y-t)^{\frac{1-2 b}{b-1}} \\ \gamma_{12}^{(1)}=\gamma_{22}^{(1)}=0\end{array}$,
 \\

&$\begin{array}{l}a_{4,16}^{a}\\ a=1*[a=\pm1]\end{array}$&$\begin{array}{c}\big\{F+\frac{1}{2}K_2;-K_1+L_3\\ +a(P_0+P_2),P_0-P_2,P_1\big\}\end{array}$& $\left(
\begin{array}{ccc}
 -c & 0 & c \\
 0 & 0 & 0 \\
 c & 0 & -c \\
\end{array}
\right)$, leads to subalgebra with 5 KVs \\

$ \begin{array}{c} \text{Extended}\\ \text{2d Poincare}\end{array}$&$a_{4,17}=b_{4,17}$ & $\left\{ F,L_3,P_1,P_2 \right\}$  & $\left(
\begin{array}{ccc}
 \frac{c}{t} & 0 & 0 \\
 0 & \frac{c}{2 t} & 0 \\
 0 & 0 & \frac{c}{2 t} \\
\end{array}
\right)$ \\
\hline 
 \end{tabular}
 \label{tablesim}
 \end{table}
\end{center}

\newpage
The first five subalgebras, we have commented in the main text.
The sixth subalgebra $a_{4,5}$ conserves $\gam$ which depends on the difference $y-t$, analogously to dependency on one coordinate if we made transformation $z\rightarrow y-t$, for $z$ a new coordinate. Redefining the translational KVs one may expect to obtain analogous subalgebra and $\gam$.

We have commented the subalgebra $a_{4,6}$ above the equation (\ref{poind}), while the subalgebra $a_{4,7}$ for constant $\gam$ agrees with $\gam$ for 5 KV subalgebras, when we add one more KV to $a_{4,7}$. The similar behaviour appears in one realization for the $a_{4,10}$ and $a_{4,11}$. Both of them contain three transitional KVs, admit constant components in $\gam$ and lead to four different $\gam$, one of which admits one further KV. Additionally, $a_{4,11}$ contains $\gam$ with imaginary value. Four subalgebras similar to $a_{4,5}$ that contain $\gam$ dependent on $y-t$  are $a^{\epsilon}_{4,12}$, $a_{4,13}$, $\tilde{a}_{4,14}$ and $a_{4,15}$.

\subsection{opt(2,1)}

The generators of optical algebra can be defined with 

\begin{align}
W&=-\frac{\xi^{(6)}+\xi^{(4)}}{2} & K_1&=\frac{\xi^{(6)}-\xi^{(4)}}{2} \\ K_2&=\frac{1}{2}\left[\xi^{(0)}-\xi^{(2)}+\frac{(\xi^{(8)}-\xi^{(9)})}{2}\right] &
L_3&=\frac{1}{2}\left[\xi^{(0)}-\xi^{(2)}-\frac{(\xi^{(8)}-\xi^{(9)})}{2}\right] \\ M&=-\sqrt{2}\xi^{(1)} & Q&=\frac{\xi^{(5)}-\xi^{(3)}}{2\sqrt2} \\
N&=-(\xi^{(0)}+\xi^{(2)}). & &
\end{align}
The generators are not uniquely defined. We can automatically see permutative dependence on KVs in definition of generators, 
and on a result with analogously permutative structure of components in $\gam$. For example, $W$ generator can be defined with $\xi^{(6)}$ dilatation and with one L. rotation. However, we can permutatively define it with $\xi^{(3)}$ or $\xi^{(5)}$ KVs of L. rotations as long as we correspondingly modify remaining KVs that define the algebra.F
The generators close the algebra
\begin{align}
[K_1,K_2]&=-L_3,  & [L_3,K_1]&=K_2, & [L_3,K_2]&=-K_1, & [M,Q]&=-N, & &,  \\ [K_1,M]&=-\frac{1}{2} M, &[K_1,Q]&=\frac{1}{2}Q, & [K_1,N]&=0, &[K_2,M]&=\frac{1}{2}Q, \\
 [K_2,Q]&=\frac{1}{2}M, & [K_2,N]&=0   &
[L_3,M]&=-\frac{1}{2}Q, & [L_3,Q]&=\frac{1}{2}M, \\  [L_3,N]&=0 &
[W,M]&=\frac{1}{2}M,  & [W,Q]&=\frac{1}{2}Q, &[W,N]&=\frac{1}{2}N.
  \end{align}
 The five realised subalgebras we write in the Table V.

\begin{center}
\begin{table}
\caption{Classfication according to \cite{Patera:1976my}, optical algebra}
\hspace{-1.2cm}\begin{tabular}{ | l  | p{2.5 cm} | p{3.9 cm} | p{5.2cm} |}
\hline
  \multicolumn{4}{|c|}{Realised subalgebras} \\
  \hline
$\begin{array}{c}Name/ \\ commutators\end{array}$&Patera name&generators  & realisation  \\ \hline\hline
 &$b_{5,6}=a_{5,4}$ & $W+aK_1,K_2+L_3,M,Q,N$  & \\
 &$b_{4,1}$&$N\oplus\{;K_1,K_2,L_3\}$ & $\left(
\begin{array}{ccc}
 \frac{c}{x^3} & 0 & -\frac{c}{x^3} \\
 0 & 0 & 0 \\
 -\frac{c}{x^3} & 0 & \frac{c}{x^3} \\
\end{array}
\right)$ \\
 &$b_{4,2}$& $W\oplus\left\{ K_1,K_2,L_3 \right\}$ & see eq. (\ref{eqn1}) \\
 &$b_{4,3}$& $L_3,Q,M,N$ & see eq. (\ref{b43}) \\
&$\begin{array}{c}b_{4,4}\\ b>0\ast [b\neq0]\end{array}$& $W+bL_3,Q,M,N$ &  see eq. (\ref{b44}).\\
$\begin{array}{c} Extended\\ 2d Poincare\end{array}$ &$\overline{b}_{4,9}=a_{4,6}$ & 
$ \begin{array}{c} \left\{K_1,K_2+L_3 \right\}\oplus  \\ \left\{W,N\right\} \end{array} $
 &  see eq. (\ref{poind}) \\ 
 &$\begin{array}{c} \overline{b}_{4,11}\sim a_{4,8}^{-1} \end{array}$ & $\begin{array}{l} \big\{ W-K_1+Q\\ K_2+L_3,M,N\big\}\end{array}$ &$\left(
\begin{array}{ccc}
 -\frac{5 \text{c}}{3 \sqrt{2}} & \text{c} & \frac{\text{c}}{\sqrt{2}} \\
 \text{c} & -\frac{1}{3} \left(2 \sqrt{2} \text{c}\right) & -\text{c} \\
 \frac{\text{c}}{\sqrt{2}} & -\text{c} & -\frac{\text{c}}{3 \sqrt{2}} \\
\end{array}
\right) $ \\

 &$\begin{array}{l}\overline{b}^b_{4,17}=a_{4,15}^{\text{\tiny{ (1-b)/(1+b)}}}\\ b>0,b\neq1\end{array}$ & $\begin{array}{l}\big\{W-bK_1,M,Q,N \big\}\end{array}$ & 
$\begin{array}{l} \gamma_{11}^{(1)}=c\cdot(y-t)^{-\frac{3 b+1}{2 b}},\\ \gamma_{13}^{(1)}=-c\cdot(y-t)^{-\frac{3 b+1}{2 b}},\\ \gamma_{33}^{(1)}=c\cdot(y-t)^{-\frac{3 b+1}{2 b}} \\ \gamma_{12}^{(1)}=\gamma_{22}^{(1)}=0\end{array}$,
\\
\hline
\end{tabular}
\end{table}
\end{center}

\noindent while $\gamma_{ij}^{(1)}$ matrices for $b_{4,2}, b_{4,3}$ and $b_{4,4}$ subalgebras, are respectively\begin{align}
\gamma_{ij}^{(1)}&=\left(
\begin{array}{ccc}
 \frac{c\cdot\left(x^2+3 (t+y)^2\right) }{x^3} & -\frac{3 c\cdot(t+y) }{x^2} & -\frac{3 c\cdot(t+y)^2 }{x^3} \\
 -\frac{3c \cdot(t+y) }{x^2} & \frac{2 c}{x} & \frac{3 c\cdot(t+y) }{x^2} \\
 -\frac{3 c\cdot(t+y)^2 }{x^3} & \frac{3 c\cdot(t+y) }{x^2} & -\frac{c\cdot\left(x^2-3 (t+y)^2\right) }{x^3} \\
\end{array}
\right)\label{eqn1},\\
\gamma_{ij}^{(1)}&=\left(
\begin{array}{ccc}
 -\frac{c}{\left((t-y)^2+4\right)^{3/2}} & 0 & \frac{c}{\left((t-y)^2+4\right)^{3/2}} \\
 0 & 0 & 0 \\
 \frac{c}{\left((t-y)^2+4\right)^{3/2}} & 0 & -\frac{c}{\left((t-y)^2+4\right)^{3/2}} \\
\end{array}
\right)\label{b43},\\
\gamma_{ij}^{(1)}&=\left(
\begin{array}{ccc}
 \frac{c\cdot e^{\frac{\tan ^{-1}\left(\frac{t-y}{2}\right)}{b}}}{\left((t-y)^2+4\right)^{3/2}} & 0 & -\frac{c\cdot e^{\frac{\tan ^{-1}\left(\frac{t-y}{2}\right)}{b}} }{\left((t-y)^2+4\right)^{3/2}} \\
 0 & 0 & 0 \\
 -\frac{c\cdot e^{\frac{\tan ^{-1}\left(\frac{t-y}{2}\right)}{b}} }{\left((t-y)^2+4\right)^{3/2}} & 0 & \frac{c \cdot e^{\frac{\tan ^{-1}\left(\frac{t-y}{2}\right)}{b}} }{\left((t-y)^2+4\right)^{3/2}} \\
\end{array}
\right)\label{b44}.
\end{align}
Here, an asterisk after the range under $o(3,2)$ as in $a^{\epsilon}_{4,12}$ means that one has to double the range in case of considering conjugacy under $SO_0(3,2)$ (or $O_1(3,1)$) rather than 
under $O(3,2)$ (or $SO(3,2)$). E.g. $\epsilon=1\ast$ implies $\epsilon=\pm1$ under $SO_0(3,2)$, while $b>0\ast$ means that $b\neq0,-\infty<b<\infty$ under $SO_0(3,2)$.

Comparing the above matrices to those from the above section "Appendix: Classification according to the generators of the conformal group" formed by the original KVs,
we can notice the similar behaviour in the components of the $\gam$ matrix of the $b_{4,1}$ group and the one that conserves two SCTs. The reason is in the SCTs that build the $K_2$ and $L_3$ of $b_{4,1}$. The $b_{4,2}$ and $b_{4,3}$ contain the power of $3/2$ in $\gam$ which also appears in the $\gam$ for 3R+D.\\
The further realised $b$ subalgebras that correspond to those of $sim(2,1)$ group are $b_{4,5}=a_{4,18}, \overline{b}_{4,6}=a_{4,1}, \overline{b}_{4,10}=a_{4,7},\overline{b}^b_{4,13}=a^{(b-1)/(b+1)}_{4,10} \text{ for }0<|b|<1 \text{ and }[b\neq0,\pm1], \overline{b}^{\epsilon}_{4,14}=a^{\epsilon}_{4,12}, \text{ here }\epsilon=1\ast[\epsilon=\pm1],\overline{b}_{4,15}=a_{4,13},\overline{b}_{4,16}=a_{4,14}$

 In $o(3)\oplus o(2)$ there are no realised subalgebras, while $o(2)\oplus o(2,1)$ contains one algebra with four generators which is formed by 1T+1SCT+1R+D and isomorphic to $sl(2)\oplus u(1)$.
 Further we consider $o(2,2)$.

\subsection{o(2,2)}
The generators of $o(2,2)$ can be written as

 \begin{align}
A_1&=-\frac{1}{2}\left[\frac{\xi^{(9)}+\xi^{(8)}}{2}-(\xi^{(0)}+\xi^{(2)})\right], && A_2=\frac{1}{2}(\xi^{(6)}+\xi^{(4)}),  \nonumber \\
 A_3&=\frac{1}{2}\left[-\frac{\xi^{(9)}+\xi^{(8)}}{2}-(\xi^{(0)}+\xi^{(2)})\right], &&
B_1=-\frac{1}{2}\left[\frac{-\xi^{(9)}+\xi^{(8)}}{2}+(\xi^{(0)}-\xi^{(2)})\right], \nonumber \\ B_2&=\frac{1}{2}(\xi^{(6)}-\xi^{(4)}),  && B_3=\frac{1}{2}\left[\frac{\xi^{(9)}-\xi^{(8)}}{2}+(\xi^{(0)}-\xi^{(2)})\right]. \label{genlo22}
\end{align}
\noindent $\gam$ dependent on the remaining two co-ordinates is obtained by permutation of the original generators. Its algebra is isomorphic to $o(2,1)\oplus o(2,1)$ 
\begin{align}
[A_1,A_2]&=-A_3,& [A_3,A_1]&=A_2, & [A_2,A_3]&=A_1, \\
[B_1,B_2]&=-B_3, & [B_3,B_1]&=B_2, & [B_2,B_3]&=B_1, \\ [A_i,B_k]&=0 &(i,k=1,2,3).   
\end{align}
Its subagebras contain from six to one generators, however algebras with 6 and 5 generators are not realised.
The subalgebras with four generators correspond to $\overline{e}_{4,2}=a_{4,6},\overline{e}_{4,3}=b_{4,1},\overline{e}_{4,4}=b_{4,2}$.

\subsection{o(3,1)}

Interesting property of the four dimensional Lorentz algebra $o(3,1)$ is that we can write it on the three dimensional hypersurface 
via 
\begin{align}
L_1&=\xi^{(7)}+\frac{\xi^{(2)}}{2}, & L_2&=\xi^{(5)}, & L_3&=\xi^{(8)}+\frac{1}{2}\xi^{(1)}, \\
K_1&=\xi^{(8)}-\frac{1}{2}\xi^{(1)}, & K_2&=\xi^{(6)}, & K_3&=-\xi^{(7)}+\frac{1}{2}\xi^{(2)} .
\end{align}
that form the commutation relations
\begin{align}
[L_i,L_j]&=\epsilon_{ijk}L_k, & [L_i,K_j]&=\epsilon_{ijk}K_k, & [K_i,K_j]&=-\epsilon_{ijk}L_k.
\end{align}

\noindent The algebra $o(3,1)$ is not realised, while its first highest sub algebra of four generators $K_1,L_1;L_2-K_3,L_3+K_2$, is equal to the  $a_{4,17}$, i.e. 2 dimensional Poincare algebra (\ref{poind}).
\\

These $\gam$ matrices for the flat background metric can be transformed into the $\gam$ on the spherical background using the map to spherical coordinates \cite{thesis}.




\bibliographystyle{abbrv}
\bibliography{bibliothek1}
 
\end{document}